\pdfoutput=1

\documentclass[11pt,twoside,a4paper,cmspaper,final,collab]{cms-tdr}
\def\svnVersion{298643}\def\cmsCernNo{2013-037}\def\cmsCernDate{\today}\def\cmsMessage{Published in the European Physical Journal C as \href{http://dx.doi.org/10.1140/epjc/s10052-015-3451-4}{\doi{10.1140/epjc/s10052-015-3451-4.}}}

\begin{document}\cmsNoteHeader{EXO-12-048}

\hyphenation{had-ron-i-za-tion}
\hyphenation{cal-or-i-me-ter}
\hyphenation{de-vices}
\RCS$Revision: 284807 $
\RCS$HeadURL: svn+ssh://svn.cern.ch/reps/tdr2/papers/EXO-12-048/trunk/EXO-12-048.tex $
\RCS$Id: EXO-12-048.tex 284807 2015-04-16 10:21:57Z smalik $

\newcommand{\LambU}{\ensuremath{{\Lambda_\cmsSymbolFace{U}}}\xspace}
\newcommand{\dU}{\ensuremath{{d_\cmsSymbolFace{U}}}\xspace}
\newcommand{\Wmunu}{\ensuremath{\PW (\mu \nu )}\xspace}
\newcommand{\Zmumu}{\ensuremath{\cPZ (\mu \mu)}\xspace}
\newcommand{\Zellell}{\ensuremath{\cPZ (\ell \ell)}\xspace}
\newcommand{\Znunu}{\ensuremath{\cPZ (\nu \bar{\nu} )}\xspace}
\newcommand{\ZnunuJets}{{\ensuremath{\cPZ (\nu \bar{\nu})}}+jets\xspace}
\newcommand{\ZellellJets}{{\ensuremath{\cPZ (\ell \ell   )}}+jets\xspace}
\newcommand{\ZJets}{{\ensuremath{\cPZ}}+jets\xspace}
\newcommand{\WJets}{{\ensuremath{\PW}}+jets\xspace}
\newcommand{\WmunuJet}{{\ensuremath{\PW (\mu \nu)}}+jets\xspace}
\newcommand{\DM}{\ensuremath{\chi}\xspace}
\newcommand{\ppbar}{\ensuremath{\Pp\Pap}\xspace}
\newcommand{\wellnubr}{\ensuremath{\PW (\ell \nu )}\xspace}
\newlength\cmsFigWidth
\ifthenelse{\boolean{cms@external}}{\setlength\cmsFigWidth{0.48\textwidth}}{\setlength\cmsFigWidth{0.48\textwidth}}
\ifthenelse{\boolean{cms@external}}{\providecommand{\cmsLeft}{Top}}{\providecommand{\cmsLeft}{Left}}
\ifthenelse{\boolean{cms@external}}{\providecommand{\cmsRight}{Bottom}}{\providecommand{\cmsRight}{Right}}

\cmsNoteHeader{EXO-12-048}
\title{\texorpdfstring{Search for dark matter, extra dimensions, and unparticles in monojet events in proton-proton collisions at $\sqrt{s} = 8$\TeV}{Search for dark matter, extra dimensions, and unparticles in monojet events in proton-proton collisions at sqrt(s) = 8 TeV}}
\titlerunning{Search for dark matter, extra dimensions, and unparticles in monojet events\ldots}

\date{\today}
\abstract{
Results are presented from a search for particle dark matter (DM), extra dimensions, and unparticles using events containing a jet and an imbalance in transverse momentum. The data were collected by the CMS detector in proton-proton collisions at the LHC and correspond to an integrated luminosity of 19.7\fbinv at a centre-of-mass energy of 8\TeV. The number of observed events is found to be consistent with the standard model prediction. Limits are placed on the DM-nucleon scattering cross section as a function of the DM particle mass for spin-dependent and spin-independent interactions. Limits are also placed on the scale parameter $M_\mathrm{D}$ in the ADD model of large extra dimensions, and on the unparticle model parameter $\LambU$. The constraints on ADD models and unparticles are the most stringent limits in this channel and those on the DM-nucleon scattering cross section are an improvement over previous collider results.}

\hypersetup{%
pdfauthor={CMS Collaboration},
pdftitle={Search for dark matter, extra dimensions, and unparticles in monojet events in proton-proton collisions at sqrt(s) = 8 TeV},%
pdfsubject={CMS},%
pdfkeywords={CMS, physics, dark matter, ADD, extradimensions, unparticle, monojet}}

\maketitle
\section{Introduction}

This paper describes a search for new physics using the signature of a hadronic jet
and an imbalance in transverse energy resulting from undetected particles. We use the term ``monojet'' to describe events
with this topology. Such events can be produced in new
physics scenarios, including particle dark matter (DM) production, large extra
dimensions, and unparticles. The data sample corresponds to an integrated
luminosity of 19.7\fbinv collected by the CMS
experiment in proton-proton collisions provided by the CERN LHC at a centre-of-mass energy of 8\TeV.

Particle dark matter has been proposed to explain numerous astrophysical
measurements, such as the rotation curves of galaxies and gravitational
lensing~\cite{{DarkMatterReview},{DMGeneral}}. Popular models of
particle dark matter hypothesize the existence of non-relativistic
particles that interact weakly with the standard model (SM) particles.
These are known as weakly interacting massive particles (WIMPs). Such models are consistent with the thermal relic abundance for dark matter~\cite{{bib:WMAP9},{bib:Planck}} if the WIMPs
have weak-scale masses and if their interaction cross section with
baryonic matter is of the order of electroweak cross sections. Some new
physics scenarios postulated to explain the hierarchy problem also
predict the existence of WIMPs~\cite{bib:SUSY}.

Since WIMPs are weakly interacting and neutral, they are not expected to produce any discernible signal in the LHC detectors. Like neutrinos,
they remain undetected and their presence in an event must be inferred
from an imbalance of the total momentum of all reconstructed particles in the plane transverse to the beam
axis. The magnitude of such an imbalance is referred to as missing transverse energy, denoted by \MET. The monojet signature can be used to search for the pair
production of WIMPs in association with a jet from initial-state radiation
(ISR), which is used to tag or trigger the event.

\begin{figure*}[!Hhtb]
  \centering
  \includegraphics[width=0.3\textwidth]{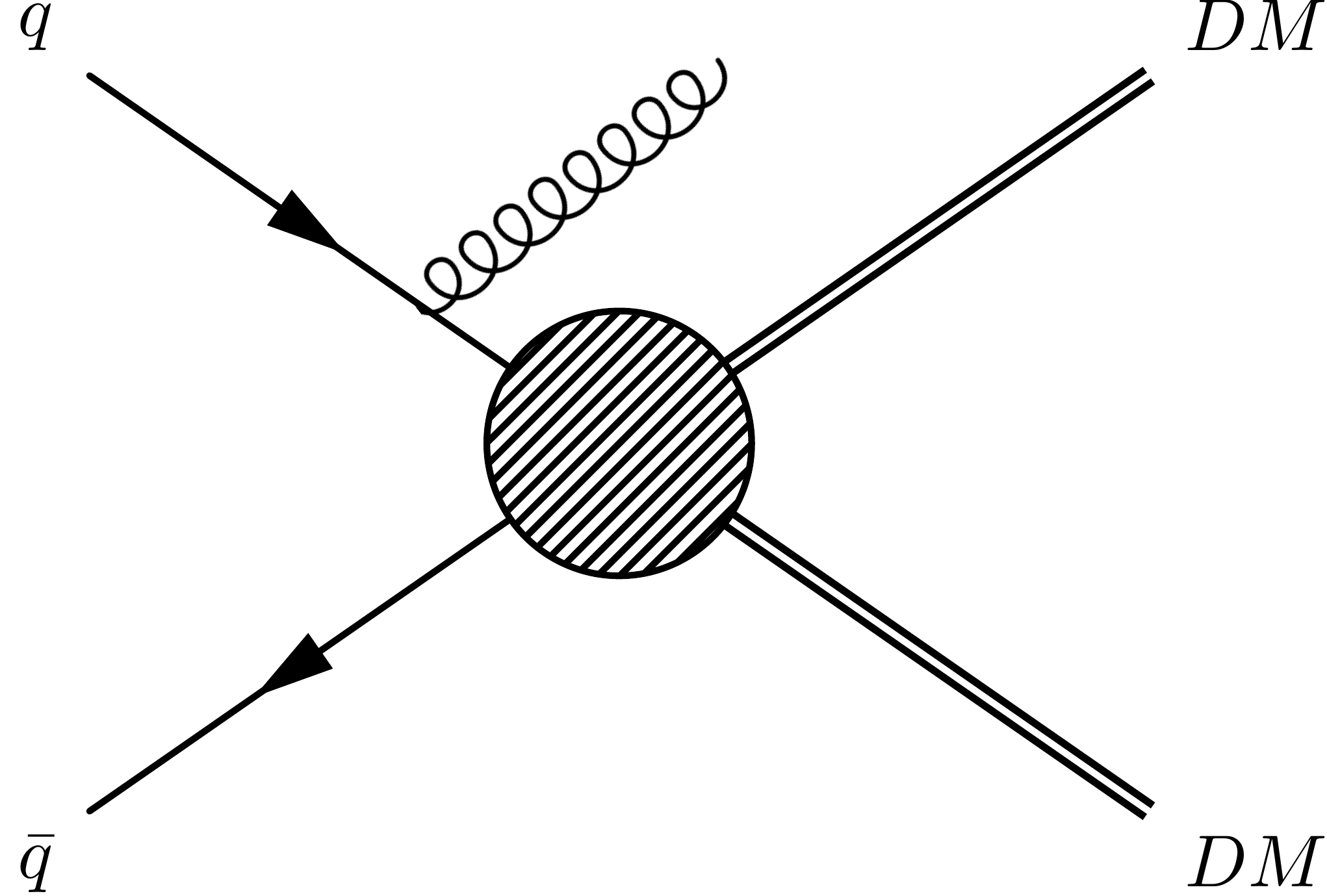}\hspace{0.15\textwidth}
  \includegraphics[width=0.3\textwidth]{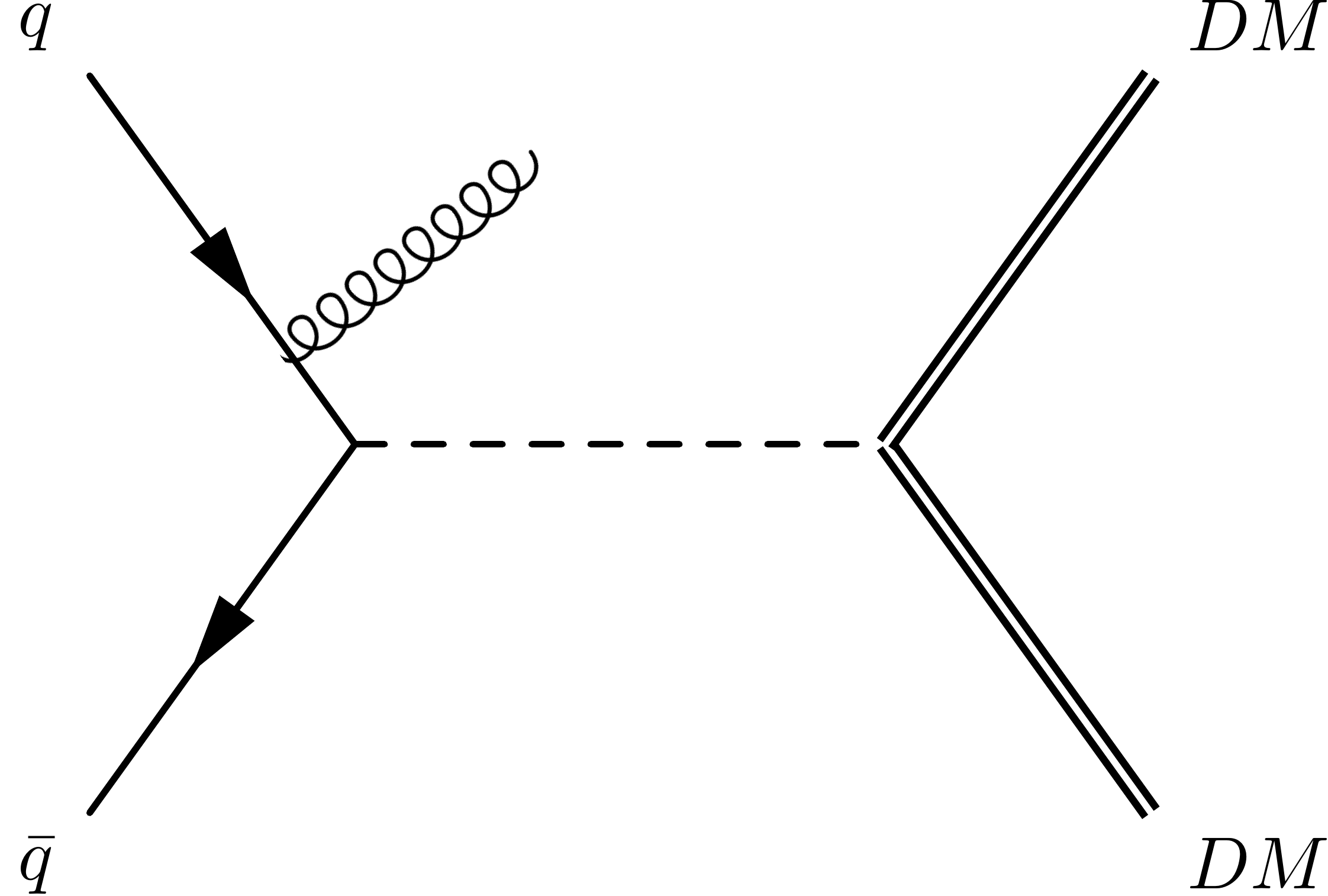}
   \caption{Feynman diagrams for the pair production of DM particles for the case of a contact interaction (left) and the exchange of a mediator (right).
         \label{fig:feynDM}}

\end{figure*}

In this Letter, we investigate two scenarios for producing dark matter particles
that have been extensively discussed~\cite{bib:TMTait,bib:TMTait2,TimTaitB,bib:RoniHarnik}. In the first case, we assume
that the mediator responsible for coupling of the SM and DM particles is heavier ($\gtrsim$ few \TeVns) than the typical energy transfer at the LHC. We can thus assume the
interaction to be a contact interaction and work within the framework of
an effective field theory. In the second case, we consider the scenario
in which the mediator is light enough to be produced at the LHC.
Figure~\ref{fig:feynDM} shows Feynman diagrams leading to the pair production of DM particles for the case of a contact interaction and the exchange of a mediator.

We study
interactions that are vector, axial-vector, and scalar, as described
in~\cite{bib:TMTait,bib:RoniHarnik}, for a Dirac fermion DM particle ($\chi$). The results are not expected to be
greatly altered if the DM particle is a Majorana fermion, except that
certain interactions are not allowed.
Results from previous searches in the monojet
channel have been used to set limits on the DM-nucleon scattering cross section as a
function of the DM mass~\cite{bib:CDFmonojet,bib:CMSEXO11059,bib:ATLAS2012ky}.

The Arkani--Hamed, Dimopoulos, and Dvali (ADD) model
\cite{bib:ADD1,ADDPRD,Antoniadis,ADDGiudice,ADDPeskin} of large extra dimensions mitigates the
hierarchy problem~\cite{Witten1981267} by introducing a number $\delta$ of extra dimensions. In
the simplest scenario, these are compactified over a multidimensional torus
with radii $R$. Gravity is free to propagate into the extra dimensions, while
SM particles and interactions are confined to ordinary space-time. The
strength of the gravitational force is thus diluted in 3+1 dimensional
space-time, explaining its apparent weakness in comparison to the other
fundamental forces. The fundamental Planck scale in $3+\delta$ spatial dimensions, \MD, is related to the apparent Planck scale in 3 dimensions, $\Mpl$ as $\Mpl^{2} = 8\pi \MD^{(\delta+2)}R^{\delta}$~\cite{ADDGiudice}.
The increased phase space available in the extra dimensions is expected to
enhance the production of gravitons, which are weakly interacting and escape undetected, their presence must therefore be inferred by detecting \MET. When
produced in association with a jet, this gives rise to the monojet
signal. Previous searches for large extra dimensions in monophoton and
monojet channels have yielded no evidence of new physics \cite{bib:OPAL,bib:ALEPH,bib:L3,bib:CDFMonoPhoton,bib:D0MonoPhoton,bib:CMS_EXO11003,bib:CMSEXO11059,bib:ATLASMonoJet,bib:ATLAS2012ky}.

\begin{figure}
  \begin{center}
  \includegraphics[scale=0.25]{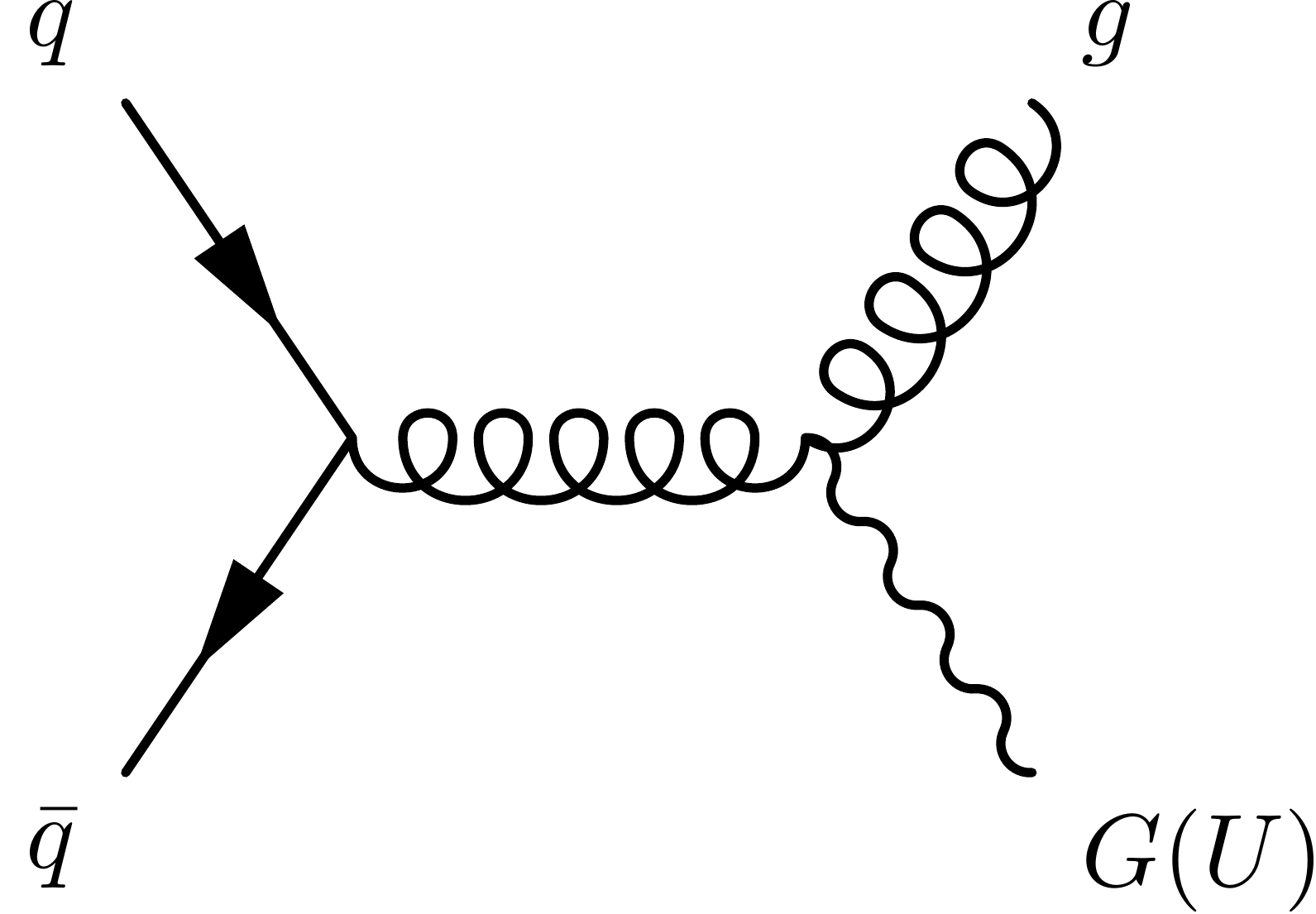}
  \includegraphics[scale=0.25]{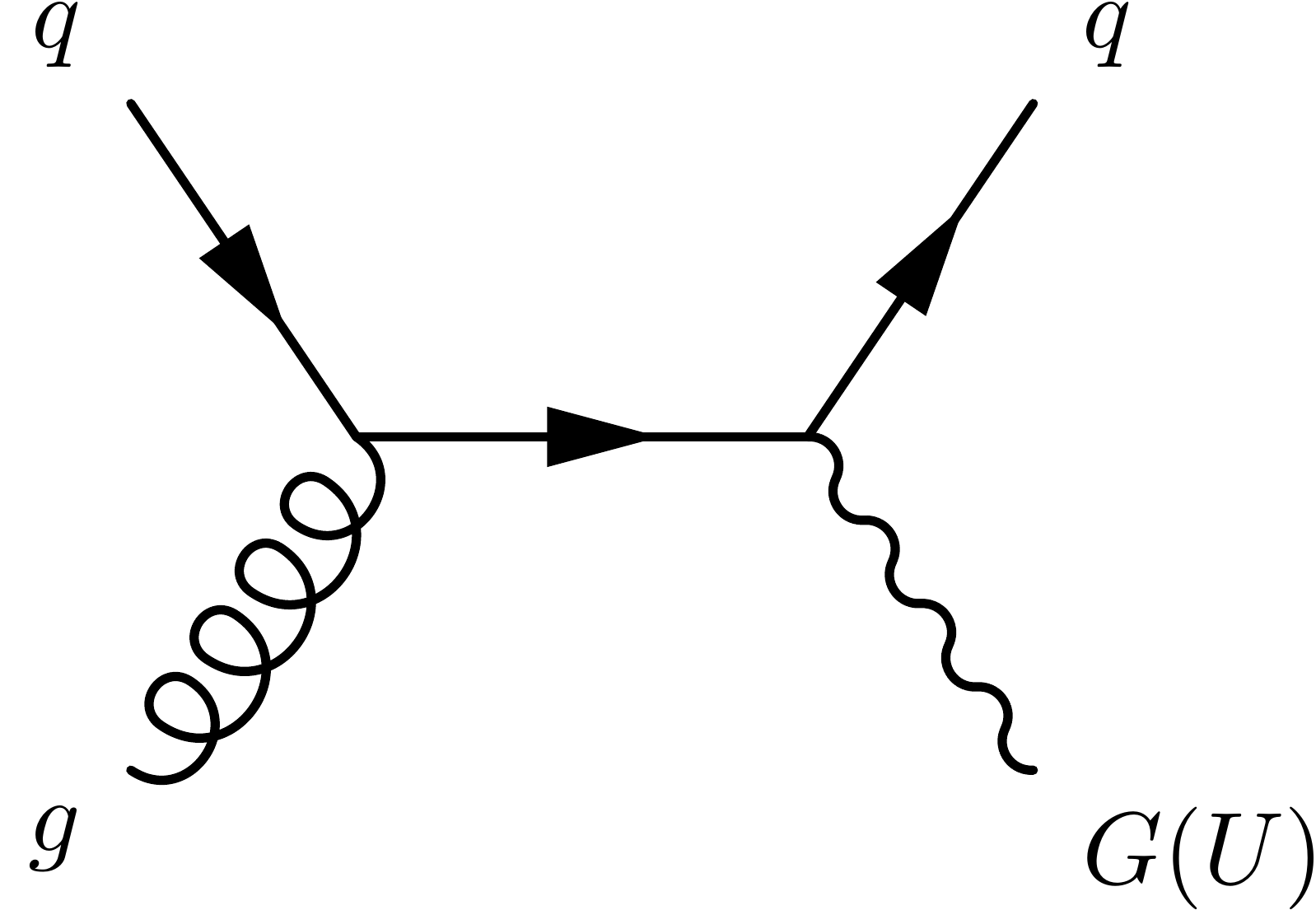}
  \includegraphics[scale=0.25]{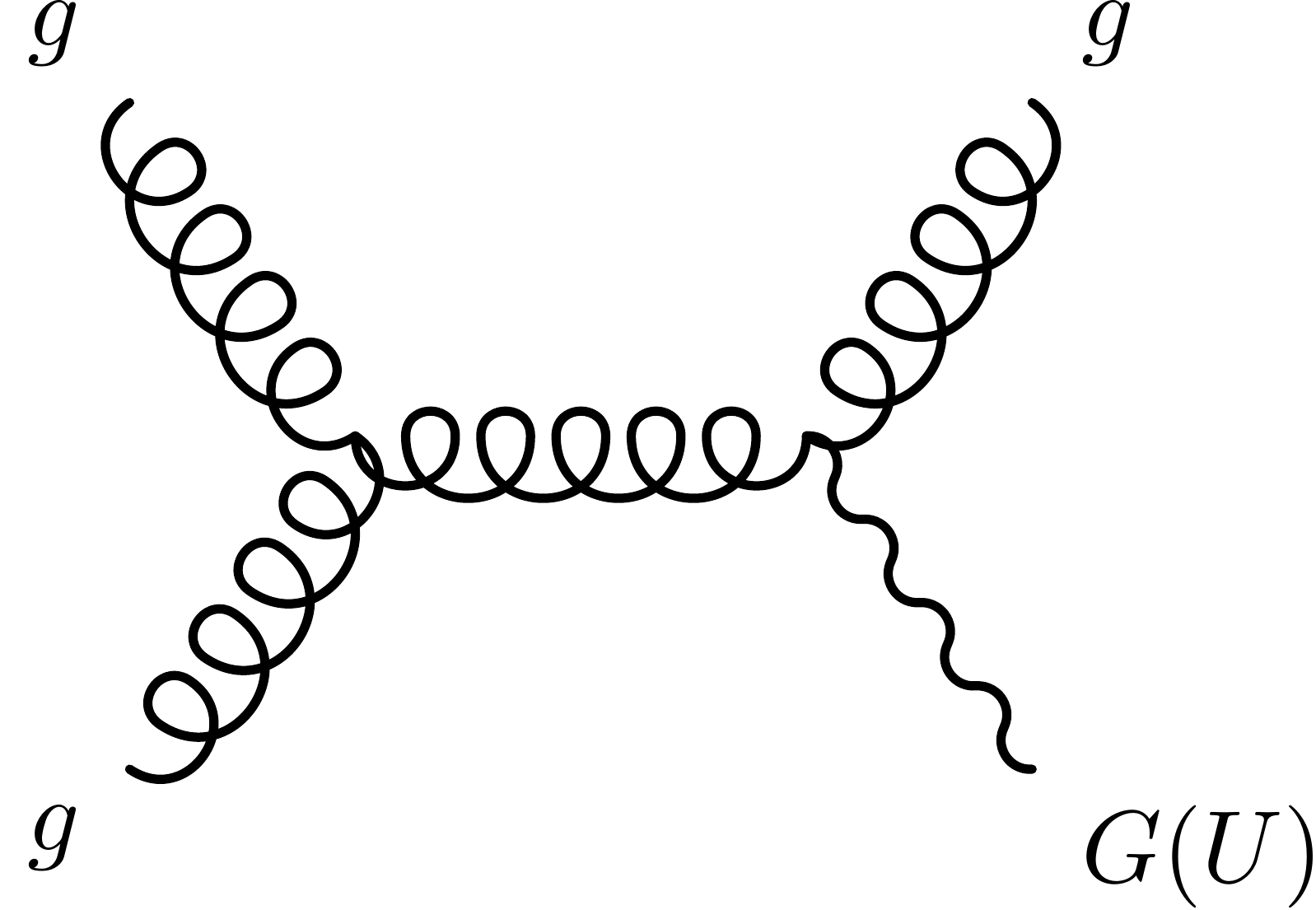}
   \caption{Feynman diagrams for the production of a graviton (G) or unparticles (U) in association with a jet.
         \label{fig:feynADD}}
  \end{center}
\end{figure}

 Unparticle models \cite{bib:Unp} postulate the existence of a
scale-invariant (conformal) sector, indicating new physics that cannot be described
using particles. This conformal sector is connected
to the SM at a high mass scale $\LambU$. In the low-energy limit,
with scale dimension $d_{u}$, events appear to correspond to the production of a non-integer number $d_{u}$
of invisible particles. Assuming these are sufficiently long-lived to
decay outside of the detector, they are undetected and so give rise to
$\MET$. If $\LambU$ is assumed to be of order\TeV, the effects of unparticles can be studied in the context of an effective field
theory at the LHC. Previous searches for unparticles at
CMS~\cite{bib:CMS_EXO11003} have yielded no evidence of new physics.
Figure~\ref{fig:feynADD} shows Feynman diagrams for some of the processes leading to the production of a graviton or unparticle in association with a jet.

\section{The CMS detector and event reconstruction}

The CMS apparatus features a superconducting solenoid, 12.5\unit{m} long
with an internal diameter of 6\unit{m}, providing a uniform magnetic field of
3.8\unit{T}. Within the field volume are a silicon pixel and strip tracker, a crystal electromagnetic calorimeter and a brass/scintillator hadron
calorimeter. The momentum resolution for reconstructed tracks in the central region is
about 1.5\% for non-isolated particles with transverse momenta (\pt) between 1 and 10\GeV and 2.8\% for isolated particles with \pt of 100\GeV.
The calorimeter system surrounds the tracker and consists of a scintillating lead tungstate
crystal electromagnetic calorimeter and a brass/scintillator hadron
calorimeter with coverage up to $\abs{\eta}=3$.
The quartz/steel forward hadron calorimeters extend the calorimetry coverage
up to $\abs{\eta}=5$.

A system of gas-ionization muon detectors embedded in the steel flux-return yoke of the
solenoid allows reconstruction and identification of muons in the
$\abs{\eta} < 2.4$ region.
Events are recorded using a two-level trigger system.
A more detailed description of the CMS detector and the trigger system can be found in~\cite{bib:CMS_TDR}.

Offline, particle candidates are individually identified using a particle-flow reconstruction~\cite{bib:ANA_PF, CMS-PAS-PFT-10-001}. This algorithm reconstructs
each particle produced in a collision by combining information from the
tracker, the calorimeters, and the muon system, and identifies them as either
a charged hadron, neutral hadron, photon, muon, or electron.
The candidate particles are then clustered into jets using the anti-\kt
algorithm~\cite{bib:ANA_AK} with a distance parameter of 0.5.
The energy resolution for jets is 15\% at \pt of 10\GeV, 8\% at \pt of 100\GeV, and 4\% at \pt of 1 \TeV~\cite{Jetresolution}.
Corrections are applied to the jet four-momenta as a function of the jet \pt and $\eta$ to account for residual effects of non-uniform detector response~\cite{JETJINST}. Contributions from multiple proton-proton collisions overlapping with the event of interest (pileup) are mitigated by discarding charged particles not associated with the primary vertex and accounting for the effects from neutral particles~\cite{Fastjet}.
The \MET in this analysis is defined as the magnitude
of the vector sum of the transverse momenta of all particles reconstructed
in the event, excluding muons.

\section{Event selection}
\label{evtsel}
Events are collected using two triggers, the first of which has an \MET
threshold of 120\GeV, where the $\MET$ is calculated using calorimeter
information only.  The second trigger requires a particle-flow jet with $\pt > 80$\GeV
and $\MET > 105$\GeV, where the $\MET$ is reconstructed using the
particle-flow algorithm and excludes muons. This definition of $\MET$ allows the control sample of $\cPZ\to\Pgm\Pgm$ events used for estimating the $\cPZ\to\Pgn\Pgn$ background to be collected from the same trigger as the signal sample. The trigger efficiencies are measured to be nearly 100\% for all signal regions. Events are required to have
a well-reconstructed primary vertex~\cite{bib:ANA_Tk}, which is defined as the one with the largest sum of $\pt^{2}$ of all the associated tracks, and is assumed to correspond to the hard scattering process.
Instrumental and beam-related backgrounds are suppressed by rejecting
events where less than 20\% of the energy of the highest \pt jet is
carried by charged hadrons, or more than 70\% of this energy is carried by
either neutral hadrons or photons. This is very effective in rejecting
non-collision backgrounds, which are found to be negligible.
The jet with the
highest transverse momentum ($\,\mathrm{j}_1$)
is required to have $\pt > 110\GeV$ and $\abs{\eta} < 2.4$.
As signal events typically contain jets from initial state
radiation, a second jet ($\,\mathrm{j}_2$)  with $\pt$ above 30\GeV and $\abs{\eta} < 4.5$
is allowed, provided the second jet is separated from the first in
azimuth ($\phi$) by less than 2.5 radians, $\Delta\phi(\mathrm{j_1},\mathrm{j_2}) < 2.5$.
This angular requirement suppresses Quantum ChromoDynamics (QCD) dijet events.  Events with
more than two jets with $\pt > 30$\GeV and $\abs{\eta} < 4.5$ are discarded, thereby significantly reducing background
from top-quark pair (\cPqt\cPaqt) and QCD multijet events. Processes producing
leptons, such as $\PW$ and $\cPZ$ production, dibosons, and top-quark
decays, are suppressed by rejecting events with well reconstructed and
isolated electrons with $\pt>10\GeV$, reconstructed muons~\cite{bib:muons} with
$\pt > 10\GeV$ and well-identified~\cite{bib:HPStaus} hadronically decaying tau leptons with $\pt > 20\GeV$ and
$\abs{\eta} < 2.3$.
Electrons and muons are considered isolated if the scalar sum of the $\pt$ of the charged hadrons, neutral hadrons and photon contributions computed in a cone of radius $\sqrt{\smash[b]{(\Delta \eta)^{2} + (\Delta \phi)^{2}}} = 0.4$ about the lepton direction, divided by the electron or muon $\pt$, is less than 0.2.
The analysis is performed in 7 inclusive regions of
$\MET$: $\MET >250,$ 300, 350, 400, 450, 500, 550\GeV.

\section{Monte Carlo event generation}
The DM signal samples are produced using the leading order (LO) matrix
element generator \MADGRAPH~\cite{bib:GEN_Mg} interfaced with \PYTHIA
6.4.26~\cite{bib:GEN_Py6} with tune Z2*~\cite{bib:Z2star} for parton
showering and hadronization, and the CTEQ~6L1~\cite{bib:SYST_CTEQ6M}
parton distribution functions (PDFs). The process of DM pair production is generated with up to two additional partons and a transverse momentum requirement of 80\GeV on the partons, with no matching to \PYTHIA. Only initial states with gluons and the four lightest quarks are considered and a universal coupling is assumed to all the quarks. The renormalization and factorization scales are set to  the sum of $\sqrt{M^{2} + \pt^{2}}$ for all produced particles, where $M$ is the mass of the particle.
For the heavy mediator case, where an effective field theory is assumed, DM particles with masses $M_\chi = 1$, 10, 100, 200, 400, 700, and
1000\GeV are generated. For the case of a light mediator, the mediator
mass, $M$, is varied from 50\GeV all the way up to 10 \TeV (to show the effect of the transition to heavy mediators) for DM particle masses of 50 and 500\GeV. Three separate samples are generated
for each value of $M$, with the width, $\Gamma$, of the mediator set to
$\Gamma = M/3$, $M/10$, or $M/8\pi$, where $M/3$ and $M/8\pi$ are taken as the extremes of a wide-width and narrow-width mediator, respectively.

The events for the ADD and unparticle models are generated with \PYTHIA
8.130~\cite{bib:GEN_PY8,bib:GEN_Ask} using tune 4C~\cite{TuneFourC} and
the CTEQ~6.6M~\cite{bib:SYST_CTEQ6M} PDFs.  This model is an effective
theory and holds only for energies well below \MD ($\LambU$) for the
graviton (unparticle).
For a parton-parton centre-of-mass energy $\sqrt{\hat{s}}>\MD$
($\LambU$), the simulated cross sections of the graviton
(unparticle) is suppressed by a factor
$\MD^4/\hat{s}^2$ ($\LambU^4/\hat{s}^2$)~\cite{bib:GEN_Ask}.
The renormalization and factorization scales are set to the geometric mean of the squared transverse mass of the outgoing particles.

The \MADGRAPH~\cite{bib:MG5,Alwall:2014hca} generator interfaced with \PYTHIA 6.4.26 and the CTEQ~6L1 PDFs is used to produce
vector bosons in association with jets ($\cPZ$+jets and $\PW$+jets), \cPqt\cPaqt,
or vector bosons in association with photons ($\PW\gamma$, $\cPZ\gamma$).
The QCD multijet and diboson ($\cPZ\cPZ$, $\PW\cPZ$, $\PW\PW$) processes
are generated with \PYTHIA 6.4.26 and CTEQ~6L1 PDFs. Single top-quark events are generated with
\POWHEG~\cite{bib:powheg,Alioli:2009je} interfaced with \PYTHIA 6.4.26 and CTEQ~6.6M PDFs. In all cases,
\PYTHIA 6.4.26 is used with the Z2* tune. All the generated signal and
background events are passed through a \GEANTfour~\cite{Agostinelli2003250,bib:GEN_GEANT4}
simulation of the CMS detector and reconstructed with the same algorithms as used for collision data. The effect of additional proton-proton interactions in each beam crossing (pileup) is modelled by superimposing minimum bias interactions (obtained using \PYTHIA with the Z2* tune) onto the hard interaction, with the multiplicity distribution of primary vertices matching the one observed in data.

\section{Background estimate}

\begin{figure}[htb]
\centering
\includegraphics[width=\cmsFigWidth]{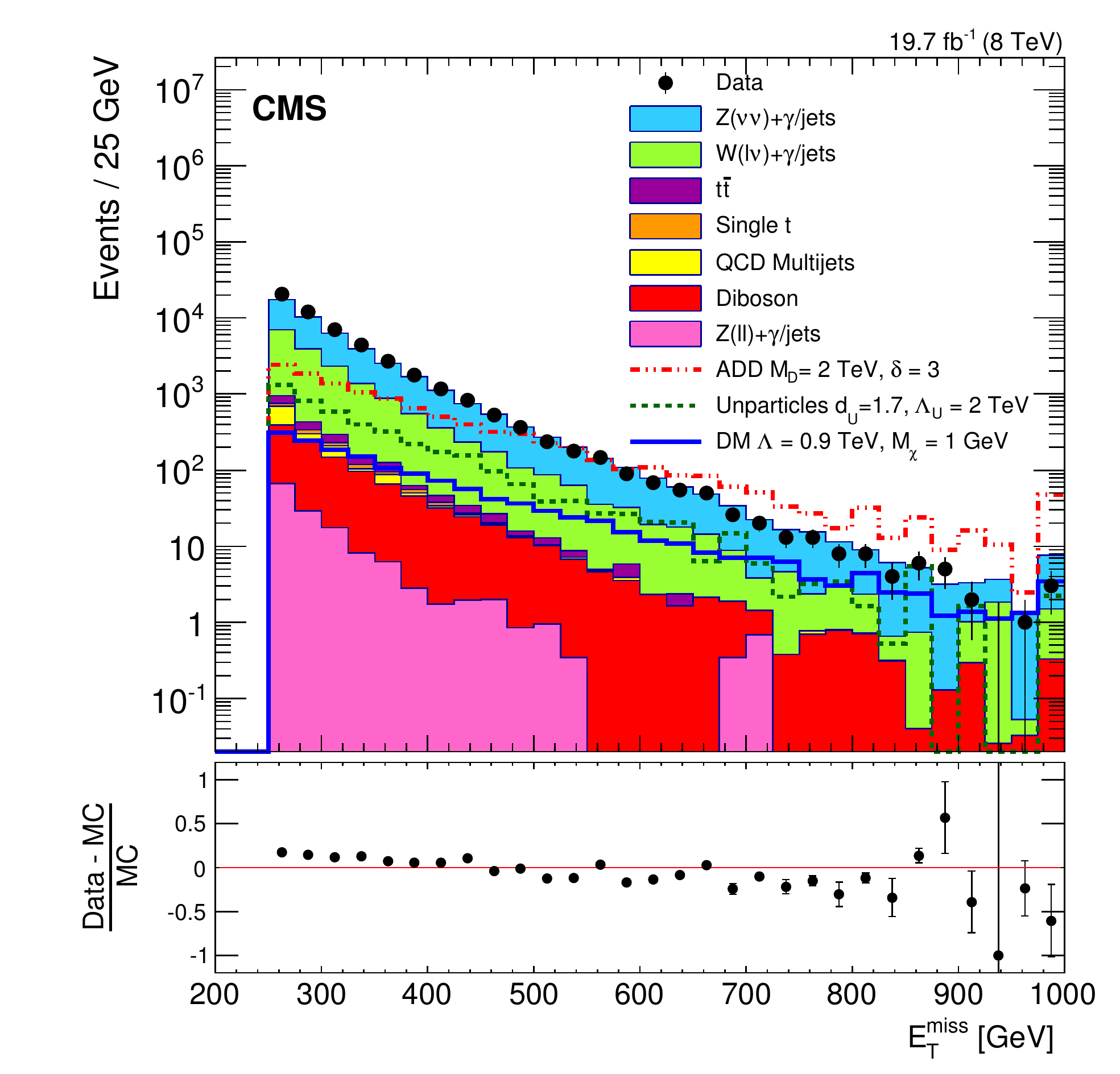}
\caption{Missing transverse energy \MET after all selections for data and SM backgrounds. The processes contributing to the SM background are from simulation, normalised to the estimation from data using the \MET threshold of 500\GeV. The error bars in the lower panel represent the statistical uncertainty. Overflow events are included in the last bin.\label{fig:ANA_MET_plots}}
\end{figure}

After the full event selection, there are two dominant backgrounds: Z+jets events with the $\cPZ$ boson decaying into a pair of neutrinos, denoted $\cPZ(\cPgn\cPgn)$; and $\PW$+jets with the $\PW$ boson decaying
leptonically, denoted $\wellnubr$ (where $\ell$ stands for a charged lepton, and can be replaced
by $\Pe$, $\Pgm$ or $\tau$ to denote specific decays to electron, muon, or tau, respectively). Other background processes include: $\cPqt\cPaqt$ production; single top quark, denoted $\cPqt$; QCD multijet; diboson processes, including $\cPZ\cPZ$, $\PW\cPZ$, and $\PW\PW$; and $\cPZ$+jets events with the $\cPZ$ boson decaying to charged leptons, denoted $\cPZ(\ell\ell)$. Together, these other background processes constitute $\approx$4\% of the total.
The dominant backgrounds are estimated from data, as described in detail below, whilst others are taken from simulation, and cross-checked with data.
Figure~\ref{fig:ANA_MET_plots} shows the $\MET$ distribution of the data and of the expected background, after imposing all the selections described in Section~\ref{evtsel} and normalised to the estimation from data using the \MET threshold of 500\GeV.

The background from events containing $\cPZ(\cPgn\cPgn)$ decays is estimated from a control
data sample of \Zmumu events, since the kinematic features of the two processes
are similar. The control sample is selected by applying the full signal selection, except for the muon veto, and in addition requiring two reconstructed muons with $\pt > 20$\GeV and $\abs{\eta} < 2.4$, with at least one muon also passing the isolation requirement. The reconstructed invariant mass is required to be between 60 and 120\GeV.
The distribution of $\cPZ(\cPgn\cPgn)$ events is estimated from the observed dimuon
control sample after correcting for the following: the estimated background in the
dimuon sample; differences in muon acceptance and efficiency with respect to neutrinos; and the
ratio of branching fractions for the $\cPZ$ decay to a pair of neutrinos, and to a pair of muons ($R_\mathrm{BF}$).  The acceptance estimate is taken from
the fraction of simulated events that pass all signal selection
requirements (except muon veto), having two generated muons with $\pt>
20$\GeV and $\abs{\eta} < 2.4$ and an invariant mass within the $\cPZ$-boson mass window of 60--120\GeV.
The efficiency of the selection, which has the additional requirement that there be at least one isolated muon in the event, is also estimated from simulation.
It is corrected to account for differences in
the measured muon reconstruction efficiencies in data and simulation.  The
uncertainty in the $\cPZ(\cPgn\cPgn)$ prediction includes both
statistical and systematic components.  The sources of uncertainty
are: (1) the statistical uncertainty in the numbers of \Zmumu events in the
data, (2) uncertainty due to backgrounds contributing to the control sample, (3) uncertainties in
the acceptance due to the size of the simulation samples and from PDFs evaluated based on the PDF4LHC~\cite{bib:PDF4LHC,bib:PDF4LHC2} recommendations, (4) the uncertainty in the selection
efficiency as determined from the difference in measured efficiencies in data and simulation and the size of the simulation samples, and (5) the theoretical uncertainty on the ratio of branching fractions~\cite{bib:BKG_PDG}.
The backgrounds to the \Zmumu control sample contribute at the level of 3--5\% across the $\MET$ signal regions and are predominantly from diboson and \ttbar processes. These are taken from simulation and a 50\% uncertainty is assigned to them.
The dominant source of uncertainty in the high $\MET$ regions is the statistical
uncertainty in the number of \Zmumu events, which is 11\% for $\MET > 500$\GeV.
Table~\ref{tab:zinv_sys} summarizes the statistical and systematic uncertainties.

\begin{table*}[htb]
        \centering
\topcaption{Summary of the statistical and systematic contributions to the total uncertainty on the $\cPZ(\cPgn\cPgn)$ background.}
\label{tab:zinv_sys}
                \begin{tabular}{l|ccccccc} \hline
\MET (\GeVns{}) $\to$& $>$250 & $>$300 & $>$350 & $>$400 & $>$450 & $>$500 &$>$550 \\ \hline
(1) $\Zmumu$+jets statistical unc. & 1.7 &  2.7 &  4.0 &  5.6 &  7.8 &  11 & 16   \\
(2) Background  & 1.4 &  1.7 &  2.1 &  2.4 &  2.7 &  3.2&   3.9\\
(3) Acceptance          & 2.0 &  2.1 &  2.1 &  2.2 &  2.3 &  2.6&   2.8\\
(4) Selection efficiency & 2.1 &  2.2 &  2.2 &  2.4 &  2.7 &  3.1 &  3.7 \\
(5) R$_\mathrm{BF}$         &    2.0 &  2.0 &  2.0 &  2.0 &  2.0 &  2.0&   2.0 \\ \hline
Total uncertainty (\%)     &    5.1 &  5.6 &  6.6 &  7.9 &  9.9 &  13 & 18 \\ \hline

\end{tabular}

\end{table*}

{\tolerance=1000 
The second-largest background arises from \WJets\ events that are not
rejected by the lepton veto. This can occur when a lepton (electron
or muon) from the W decays (prompt or via leptonic tau decay) fails the identification, isolation or acceptance requirements, or a hadronic tau decay is not identified.
The contributions to the signal region from these events are estimated
from the $\Wmunu+$jets control sample in data.  This sample is selected by applying the full signal selection, except the muon veto, and instead requiring an isolated muon with $\pt > 20$\GeV and $\abs{\eta} < 2.4$, and
the transverse mass $M_{\rm T}$ to be between 50 and 100\GeV.  Here
$M_\mathrm{T}=\sqrt{2\pt^{\mu}\MET\left(1-\cos\Delta\phi\right)}$, where
$\pt^{\Pgm}$ is the transverse momentum of the muon and $\Delta\phi$
is the azimuthal angle between the muon direction of flight and the negative of the sum of the transverse momenta of all the particles reconstructed in the event.

The observed number of events in the $\PW$ control sample is
used to find the numbers of $\Wmunu+$jets events passing the selection
steps prior to the lepton veto.  The required corrections for
background contamination of the control sample, and for the acceptance and
efficiency are taken from simulation.  Using these correction factors, we estimate the fraction
of events containing muons that are not identified, either due
to inefficiencies in the reconstruction or because they have trajectories outside
the muon system acceptance. This acceptance and the selection efficiency are also taken from simulation.  Such events will not be rejected by the
lepton veto and so contribute to the background in the signal region.

In addition, there are similar contributions from $\PW$ decays to electrons
and tau leptons. These contributions are also estimated based on the
$\Wmunu+$jets sample. The ratio of $\wellnubr+$jets events to
$\Wmunu+$jets events passing the selection steps prior to the lepton
veto is taken from simulation, separately for each lepton flavor.
The same procedure as that used in the muon case is then applied to obtain the
background contribution to the signal region.

The detector acceptances for electrons, muons and tau leptons are obtained from
simulation. The lepton selection efficiency is also obtained from simulation, but
corrected for any difference between the efficiency measured in data and
simulation~\cite{bib:tagprobe}. A systematic uncertainty of 50\% is assigned to the correction for contamination from background events taken from simulation.

The sources of uncertainty in the \WJets estimation are: (1) the statistical uncertainty in the number of single-muon events in the data, (2) uncertainty in the background events obtained from simulation, (3) uncertainty in acceptance from PDFs and size of the simulation samples and uncertainty in the selection efficiency from the variation in the data/MC scale factor and size of the simulation samples. A summary of the fractional contributions of these uncertainties to the total uncertainty in the \WJets background is shown in Table~\ref{tab:wjetssys}.

\begin{table*}[htb]
        \centering
\topcaption{Summary of the statistical and systematic contributions to the total uncertainty on the W+jets background from the various factors used in the estimation from data.}
\label{tab:wjetssys}
                \begin{tabular}{l|ccccccc} \hline
\MET (\GeVns) $\to$& $>$250 & $>$300 & $>$350 & $>$400 & $>$450 & $>$500 &$>$550 \\ \hline
(1) $\Wmunu$+jets statistical unc.  & 0.8 & 1.3 & 1.9 & 2.8 & 3.9 & 5.5 & 7.3 \\
(2) Background   & 2.3 & 2.3 & 2.2 & 2.3 & 2.4 & 2.6 & 2.8 \\
(3) Acceptance and efficiency          & 4.5 & 4.6 & 4.9 & 5.2 &5.7 & 6.4 & 7.6 \\ \hline
Total uncertainty (\%) & 5.1 & 5.3 & 5.7 & 6.4 & 7.3 & 8.8 & 11 \\ \hline
\end{tabular}
\end{table*}

The QCD multijet background is estimated by correcting the prediction from
simulation with a data/MC scale factor derived from a QCD-enriched region in data. The QCD-enriched region is selected by applying the signal selection but relaxing the requirement on the jet multiplicity and the angular separation between the first and second jet and instead requiring that the azimuth angle between the $\MET$ and the second jet is less than 0.3.
The \pt threshold for selecting jets (all except the leading jet) is varied from 20\GeV to 80\GeV and an average scale factor is derived from a comparison between data and simulation.
The {\ttbar} background is determined from simulation and normalised to the approximate next-to-next-to-leading-order cross section~\cite{Kidonakis:2010dk}, and is validated using a control
sample of $\Pe\mu$ events in data. The predictions for the number of diboson
($\PW\PW$, $\PW\cPZ$, $\cPZ\cPZ$) events are also determined from simulation,
and normalised to their next-to-leading-order (NLO) cross sections~\cite{MCFM:diboson}. Predictions
for $\PW\cPgg$ and $\cPZ(\cPgn\cPgn)\cPgg$ events are included in the estimation of {\WJets} and {$\cPZ(\cPgn\cPgn)$+jets} from data, as photons are not explicitly vetoed in the estimation of the {\WJets} and {$\cPZ(\cPgn\cPgn)$+jets} backgrounds.
Single top and {\ZellellJets} (including
$\cPZ(\ell\ell)\cPgg$ production) are predicted to contribute $\sim$0.3\% of
the total background, and are determined from simulation.  A $50\%$ uncertainty is assigned to these backgrounds. In addition to this 50\% uncertainty, the uncertainty on the QCD background also receives a contribution of 30\% arising from the uncertainty on the data/MC scale factor.

\section{Results}
A summary of the predictions and corresponding uncertainties for all the
SM backgrounds and the data is shown in Table~\ref{tab:summary_bgd} for
different values of the $\MET$ selection. The observed number of events is
consistent with the background expectation, given the statistical and
systematic uncertainties. The CL$_\mathrm{s}$ method~\cite{bib:CLs1,bib:CLs2,bib:STAT_RooStats} is employed for
calculating the upper limits on the signal cross section using a profile likelihood ratio as the test-statistic
and systematic uncertainties modeled by log-normal distributions. Uncertainties in the signal acceptance (described below) are taken into account when upper limits on the cross section are determined.
The expected and observed 95\%
confidence level (CL) upper limits on the contribution of events from new physics are also shown.
The model-independent upper limits on the visible cross section for non-SM production of events (denoted $\sigma_\text{vis}^\mathrm{BSM}$) are shown in Fig.~\ref{fig:modelindep}.

\begin{table*}[htb]
\centering
\topcaption{SM background predictions for the numbers of events passing
the selection requirements, for various \MET thresholds, compared with
the observed numbers of events.  The uncertainties include both
statistical and systematic components. The last two rows give the
expected and observed upper limits, at 95\% CL, for the contribution of
events from non-SM sources passing the selection requirements.}
\label{tab:summary_bgd}
\resizebox{\textwidth}{!}{
\begin{tabular}{l|ccccccc}
\hline
\multicolumn{1}{l|}{\MET (\GeVns{}) $\to$}&
$>$250 &
$>$300 &
$>$350 &
$>$400 &
$>$450 &
$>$500 &
$>$550 \\
\hline
{$\cPZ(\cPgn\cPgn)$+jets}        & 32100 $\pm$ 1600 & 12700 $\pm$ 720 & 5450 $\pm$ 360 & 2740 $\pm$ 220 & 1460 $\pm$ 140 & 747 $\pm$ 96 & 362 $\pm$ 64 \\
{\WJets}      &17600 $\pm$ 900 & 6060 $\pm$ 320 & 2380 $\pm$ 130 & 1030 $\pm$ 65 & 501 $\pm$ 36 & 249 $\pm$ 22 & 123 $\pm$ 13 \\
{\ttbar}        & 446 $\pm$ 220 & 167 $\pm$ 84 & 69 $\pm$ 35 & 31 $\pm$ 16 & 15 $\pm$ 7.7 & 6.6 $\pm$ 3.3 & 2.8 $\pm$ 1.4\\
{\ZellellJets}        & 139 $\pm$ 70 & 44 $\pm$ 22 & 18 $\pm$ 9.0 & 8.9 $\pm$ 4.4 & 5.2 $\pm$ 2.6 & 2.3 $\pm$ 1.2 & 1.0 $\pm$ 0.5 \\
{Single t}      & 155 $\pm$ 77 & 53 $\pm$ 26 & 18 $\pm$ 9.1 & 6.1 $\pm$ 3.1 & 0.9 $\pm$ 0.4 & --- & --- \\
{QCD multijets}    & 443 $\pm$ 270 & 94 $\pm$ 57 & 29 $\pm$ 18 & 4.9 $\pm$ 3.0 & 2.0 $\pm$ 1.2 & 1.0 $\pm$ 0.6 & 0.5 $\pm$ 0.3 \\
{Diboson}       & 980 $\pm$ 490 & 440 $\pm$ 220 & 220 $\pm$ 110 & 118 $\pm$ 59 & 65 $\pm$ 33 & 36 $\pm$ 18 & 20 $\pm$ 10 \\ \hline
{Total SM}    & 51800 $\pm$ 2000 & 19600 $\pm$ 830 & 8190 $\pm$ 400 & 3930 $\pm$ 230 & 2050 $\pm$ 150 & 1040 $\pm$ 100 & 509 $\pm$ 66 \\
{Data} & 52200 & 19800 & 8320 & 3830 & 1830 & 934 & 519 \\
\hline
{Exp. upper limit${+}1\sigma$} &
5940 &
2470 &
1200 &
639 &
410 &
221 &
187 \\
{Exp. upper limit ${-}1\sigma$} &
2870 &
1270 &
638 &
357 &
168 &
123 &
104 \\
{Exp. upper limit}    &
4250 &
1800 &
910 &
452 &
266 &
173 &
137 \\
{Obs. upper limit}    &
4510 &
1940 &
961 &
397 &
154 &
120 &
142 \\
\hline
\end{tabular}
}

\end{table*}

\begin{figure}[!Hhtb]
  \centering
  \includegraphics[width=\cmsFigWidth]{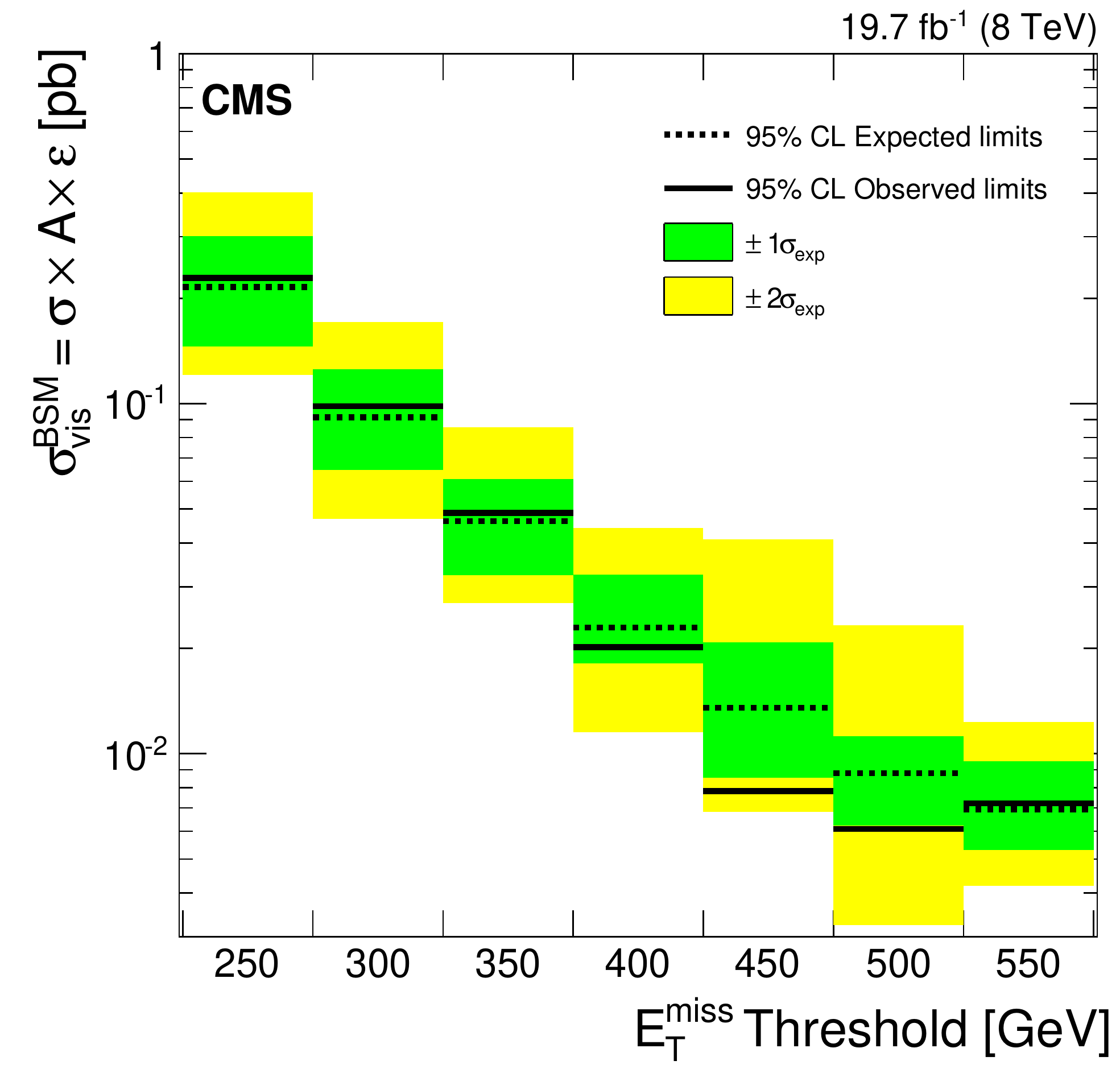}
  \caption{The model-independent observed and expected 95\% CL upper limits on the visible cross section times acceptance times efficiency ($\sigma \times A \times \varepsilon$) for non-SM production of events.  Shaded areas show the
$\pm 1\sigma$ and $\pm 2\sigma$ bands on the expected limits.}
  \label{fig:modelindep}

\end{figure}

{\tolerance=1000 
The total systematic uncertainty in the signal yield is found to be approximately 20\% for
the vector and axial-vector dark matter models, ADD extra dimensions, and unparticles, and between 20\% to 35\% for the scalar dark matter model. The sources of systematic
uncertainties considered are: jet energy scale, which is estimated by shifting the four-vectors of the jets by an $\eta$- and $\pt$-dependent factor~\cite{JETJINST}; PDFs, evaluated using the PDF4LHC prescription from the envelope of the CT10~\cite{bib:CT10}, MSTW2008NLO~\cite{bib:MSTW2008}, NNPDF2.1~\cite{bib:NNPDF} error sets;
renormalization/factorization scales, evaluated by varying simultaneously the renormalization/factorization scale up and down by a factor of 2; modeling of the ISR; simulation of event
pileup; and the integrated luminosity measurement. The PDF uncertainty is also evaluated using the LO PDFs (MSTW2008LO~\cite{bib:MSTW2008} and NNPDF21LO~\cite{bib:NNPDF}) and found to be consistent with the results from the NLO PDFs. The ISR uncertainty is estimated by varying parton shower parameters within \PYTHIA for all signal models. In addition, for the dark matter models, a further uncertainty in ISR is obtained by considering the difference in acceptance and cross section from the nominal generated samples to those where a \pt threshold of 15\GeV is applied on the generated partons and the MLM matching prescription is used to match the matrix element calculation to the parton shower in \PYTHIA, with the matching \pt scale of 20\GeV.
The dominant uncertainties are from
the modeling of the ISR, which contributes at the level of 5\% for the dark matter models
and 12\% for ADD/unparticle models, and the choice of renormalization/factorization scale,
which leads to an uncertainty of around 10\% for ADD/unparticle models and 15\% for the
dark matter models. In addition, the uncertainty on the scalar dark matter model is dominated by the PDF uncertainty, which ranges from 7\% for low DM mass and up to 30\% for high DM mass.

For each signal point, limits are derived from the signal region expected to give the best limit on the cross section.
For dark matter and ADD models, the most stringent limits are obtained for $\MET > 500$\GeV, whereas for unparticles the optimal selection varies from $\MET > 300$\GeV for $\LambU = 1$\TeV to $\MET > 500$\GeV for larger values of $\LambU$.

\section{Interpretation}

The observed limit on the cross section depends on the mass of the dark matter particle and the nature of its interaction with the SM particles. The limits on the effective contact interaction scale $\Lambda$ as a function of $M_{\chi}$ can be translated into a limit on the dark matter-nucleon scattering cross section using the reduced mass of the $\chi$-nucleon system~\cite{bib:RoniHarnik}.

Within the framework of the effective field theory, we extract limits
 on the contact interaction scale, $\Lambda$, and on the
DM-nucleon scattering cross-section, $\sigma_{\DM \mathrm{N}}$. The confidence level chosen for these limits is 90\%, to enable a direct comparison with the results from the direct detection experiments. The expected and
observed limits as a function of the DM mass, $M_{\chi}$, are shown for the vector and axial-vector operators~\cite{bib:TMTait,bib:RoniHarnik} in Tables~\ref{tab:DM_limits_V} and~\ref{tab:DM_limits_AV}, respectively, and for the scalar operator~\cite{bib:TMTait,bib:RoniHarnik} in
Table~\ref{tab:DM_limits_S}. Figure~\ref{fig:DM_limits} shows the 90\% CL upper limits on the DM-nucleon scattering cross section as a function of $M_{\chi}$ together with those from the direct detection experiments and the previously published CMS result. The limits for the axial-vector operator translate to spin dependent interactions of the dark matter with nucleons, and for the vector and scalar operators they translate to spin independent dark matter-nucleon interactions.

Given the high centre-of-mass energies that are being probed by the LHC, it is important to consider the possibility that the effective theory is
not always valid. The validity of the effective theory has been discussed in~\cite{bib:RoniHarnik,bib:RoniLHC,bib:TMTait2,An:2012va,Friedland:2011za,Buchmueller:2013dya}. It is pointed out in the literature that for theories to be perturbative the product of the couplings $g_{\chi}g_{\Pq}$ is typically required to be smaller than 4$\pi$, and this condition is likely not satisfied for the entire region of phase space probed by the collider searches. In addition, the range of values for the couplings being probed within the effective field theory may be unrealistically large~\cite{Buchmueller:2013dya}.

Therefore, we also consider the explicit case of an $s$-channel mediator with vector interactions, following the model described in \cite{bib:RoniLHC}. The mass of the mediator is varied for two fixed values of the mass of the DM particle, 50 and 500\GeV.  The width of the mediator is varied between the extremes of $M$/$8\pi$ and $M/3$, where $M/8\pi$ corresponds to a mediator that can annihilate into only one quark flavor and helicity, has couplings $g_{\chi}g_{\Pq} = 1$ and is regarded as a lower limit on the mediator width. However, not all widths may be physically realizable for the
DM couplings that are considered~\cite{bib:RoniLHC}.
Figure~\ref{fig:DM_limits_scan} shows the resulting observed limits on
the mediator mass divided by coupling ($M/\!\sqrt{g_{\chi}g_{\Pq}}$), as
a function of the mass of the mediator.
The resonant enhancement in the
production cross section, once the mass of the mediator is within the
kinematic range and can be produced on-shell, can be clearly seen. The limits on
$M/\!\sqrt{g_{\chi}g_{\Pq}}$
approximate to those obtained from the effective field theory framework
at large mediator mass, but are weaker at low mediator mass.
Also shown are dashed contours corresponding to constant values of the couplings $g_{\chi}g_{\Pq}$.

\begin{table*}[!Hhtb]
        \centering
        \topcaption{
Expected and observed 90\% CL upper limits on the DM-nucleon cross
section, $\sigma_{\DM \mathrm{N}}$, and 90\% CL lower limits on the effective contact interaction
scale, $\Lambda$, for the vector operator.
\label{tab:DM_limits_V}}
\small
\begin{tabular}{c|cc|cc|cc|cc}\hline
$M_{\chi}$
& \multicolumn{2}{c|}{Expected}
& \multicolumn{2}{c|}{Expected ${-}1\sigma$}
& \multicolumn{2}{c|}{Expected ${+}1\sigma$}
& \multicolumn{2}{c}{Observed} \\
&  $\Lambda$
& $\sigma_{\DM \mathrm{N}}$
&  $\Lambda$
& $\sigma_{\DM \mathrm{N}}$
&  $\Lambda$
& $\sigma_{\DM \mathrm{N}}$
&  $\Lambda$
& $\sigma_{\DM \mathrm{N}}$ \\
(\!\GeV)
& (\GeVns)
& (cm$^{2}$)
& (\GeVns)
& (cm$^{2}$)
& (\GeVns)
& (cm$^{2}$)
& (\GeVns)
& (cm$^{2}$)\\ \hline
1    & 951 & $3.19{\times}10^{-40}$ & 1040 & $2.23{\times}10^{-40}$ & 843 & $5.17{\times}10^{-40}$ & 1029 & $2.33{\times}10^{-40}$  \\
10   & 959 & $9.68{\times}10^{-40}$ & 1049 & $6.77{\times}10^{-40}$ & 850 & $1.57{\times}10^{-39}$ & 1038 & $7.06{\times}10^{-40}$  \\
100  & 960 & $1.13{\times}10^{-39}$ & 1050 & $7.92{\times}10^{-40}$ & 851 & $1.83{\times}10^{-39}$ & 1039 & $8.26{\times}10^{-40}$  \\
200  & 926 & $1.32{\times}10^{-39}$ & 1013 & $9.21{\times}10^{-40}$ & 821 & $2.13{\times}10^{-39}$ & 1003 & $9.60{\times}10^{-40}$  \\
400  & 848 & $1.89{\times}10^{-39}$ & 927  & $1.32{\times}10^{-39}$ & 752 & $3.06{\times}10^{-39}$ & 918  & $1.37{\times}10^{-39}$  \\
700  & 652 & $5.40{\times}10^{-39}$ & 713  & $3.78{\times}10^{-39}$ & 578 & $8.75{\times}10^{-39}$ & 706  & $3.94{\times}10^{-39}$  \\
1000 & 471 & $1.99{\times}10^{-38}$ & 515  & $1.39{\times}10^{-38}$ & 418 & $3.22{\times}10^{-38}$ & 510  & $1.45{\times}10^{-38}$  \\
\hline
\end{tabular}

\end{table*}

\begin{table*}[!Hhtb]
        \centering
        \topcaption{
Expected and observed 90\% CL upper limits on the DM-nucleon cross
section, $\sigma_{\DM \mathrm{N}}$, and 90\% CL lower limits on the effective contact interaction
scale, $\Lambda$, for the axial-vector operator.
\label{tab:DM_limits_AV}}
\small
\begin{tabular}{c|cc|cc|cc|cc}\hline
$M_{\chi}$
& \multicolumn{2}{c|}{Expected}
& \multicolumn{2}{c|}{Expected ${-}1\sigma$}
& \multicolumn{2}{c|}{Expected ${+}1\sigma$}
& \multicolumn{2}{c}{Observed}  \\
&  $\Lambda$
& $\sigma_{\DM \mathrm{N}}$
&  $\Lambda$
& $\sigma_{\DM \mathrm{N}}$
&  $\Lambda$
& $\sigma_{\DM \mathrm{N}}$
&  $\Lambda$
& $\sigma_{\DM \mathrm{N}}$ \\
(\GeVns{})
& (\GeVns{})
& (cm$^{2}$)
& (\GeVns{})
& (cm$^{2}$)
& (\GeVns{})
& (cm$^{2}$)
& (\GeVns{})
& (cm$^{2}$)\\ \hline
1    & 947 & $1.19{\times}10^{-41}$ & 1035 & $8.33{\times}10^{-42}$ & 839 & $1.93{\times}10^{-41}$ & 1025 & $8.68{\times}10^{-42}$ \\
10   & 949 & $3.71{\times}10^{-41}$ & 1038 & $2.59{\times}10^{-41}$ & 841 & $6.00{\times}10^{-41}$ & 1027 & $2.70{\times}10^{-41}$ \\
100  & 932 & $4.68{\times}10^{-41}$ & 1019 & $3.28{\times}10^{-41}$ & 826 & $7.58{\times}10^{-41}$ & 1008 & $3.41{\times}10^{-41}$ \\
200  & 880 & $5.94{\times}10^{-41}$ & 962  & $4.15{\times}10^{-41}$ & 780 & $9.62{\times}10^{-41}$ & 952  & $4.33{\times}10^{-41}$  \\
400  & 722 & $1.32{\times}10^{-40}$ & 789  & $9.21{\times}10^{-41}$ & 640 & $2.13{\times}10^{-40}$ & 781  & $9.60{\times}10^{-41}$  \\
700  & 505 & $5.52{\times}10^{-40}$ & 552  & $3.86{\times}10^{-40}$ & 447 & $8.94{\times}10^{-40}$ & 546  & $4.03{\times}10^{-40}$  \\
1000 & 335 & $2.85{\times}10^{-39}$ & 366  & $1.99{\times}10^{-39}$ & 297 & $4.61{\times}10^{-39}$ & 363  & $2.08{\times}10^{-39}$  \\
\hline
\end{tabular}

\end{table*}

\begin{table*}[!Hhtb]
        \centering
        \topcaption{
Expected and observed 90\% CL upper limits on the DM-nucleon cross
section, $\sigma_{\DM \mathrm{N}}$, and 90\% CL lower limits on the effective contact interaction
scale, $\Lambda$, for the scalar operator.
\label{tab:DM_limits_S}}
\begin{tabular}{c|cc|cc|cc|cc}\hline
$M_{\chi}$
& \multicolumn{2}{c|}{Expected}
& \multicolumn{2}{c|}{Expected ${-}1\sigma$}
& \multicolumn{2}{c|}{Expected ${+}1\sigma$}
& \multicolumn{2}{c }{Observed} \\
&  $\Lambda$
& $\sigma_{\DM \mathrm{N}}$
&  $\Lambda$
& $\sigma_{\DM \mathrm{N}}$
&  $\Lambda$
& $\sigma_{\DM \mathrm{N}}$
&  $\Lambda$
& $\sigma_{\DM \mathrm{N}}$ \\
(\GeVns{})
& (\GeVns)
& (cm$^{2}$)
& (\GeVns{})
& (cm$^{2}$)
& (\GeVns{})
& (cm$^{2}$)
& (\GeVns{})
& (cm$^{2}$) \\ \hline
1    & 411 & $1.85\times 10^{-45}$ & 437 & $1.30\times 10^{-45}$ & 380 & $3.00\times 10^{-45}$ & 436 & $1.31\times 10^{-45}$ \\
10   & 407 & $6.15\times 10^{-45}$ & 432 & $4.31\times 10^{-45}$ & 375 & $1.02\times 10^{-44}$ & 430 & $4.44\times 10^{-45}$ \\
100  & 407 & $7.25\times 10^{-45}$ & 432 & $5.08\times 10^{-45}$ & 375 & $1.20\times 10^{-44}$ & 430 & $5.23\times 10^{-45}$ \\
200  & 402 & $7.96\times 10^{-45}$ & 426 & $5.58\times 10^{-45}$ & 369 & $1.31\times 10^{-44}$ & 424 & $5.75\times 10^{-45}$ \\
400  & 348 & $1.90\times 10^{-44}$ & 368 & $1.34\times 10^{-44}$ & 319 & $3.16\times 10^{-44}$ & 366 & $1.39\times 10^{-44}$ \\
700  & 274 & $7.91\times 10^{-44}$ & 290 & $5.60\times 10^{-44}$ & 252 & $1.32\times 10^{-43}$ & 289 & $5.79\times 10^{-44}$ \\
1000 & 208 & $4.15\times 10^{-43}$ & 220 & $2.94\times 10^{-43}$ & 191 & $6.93\times 10^{-43}$ & 219 & $3.04\times 10^{-43}$ \\
 \hline
\end{tabular}

\end{table*}

\begin{figure}[!Hhtb]
  \centering
  \includegraphics[width=\cmsFigWidth]{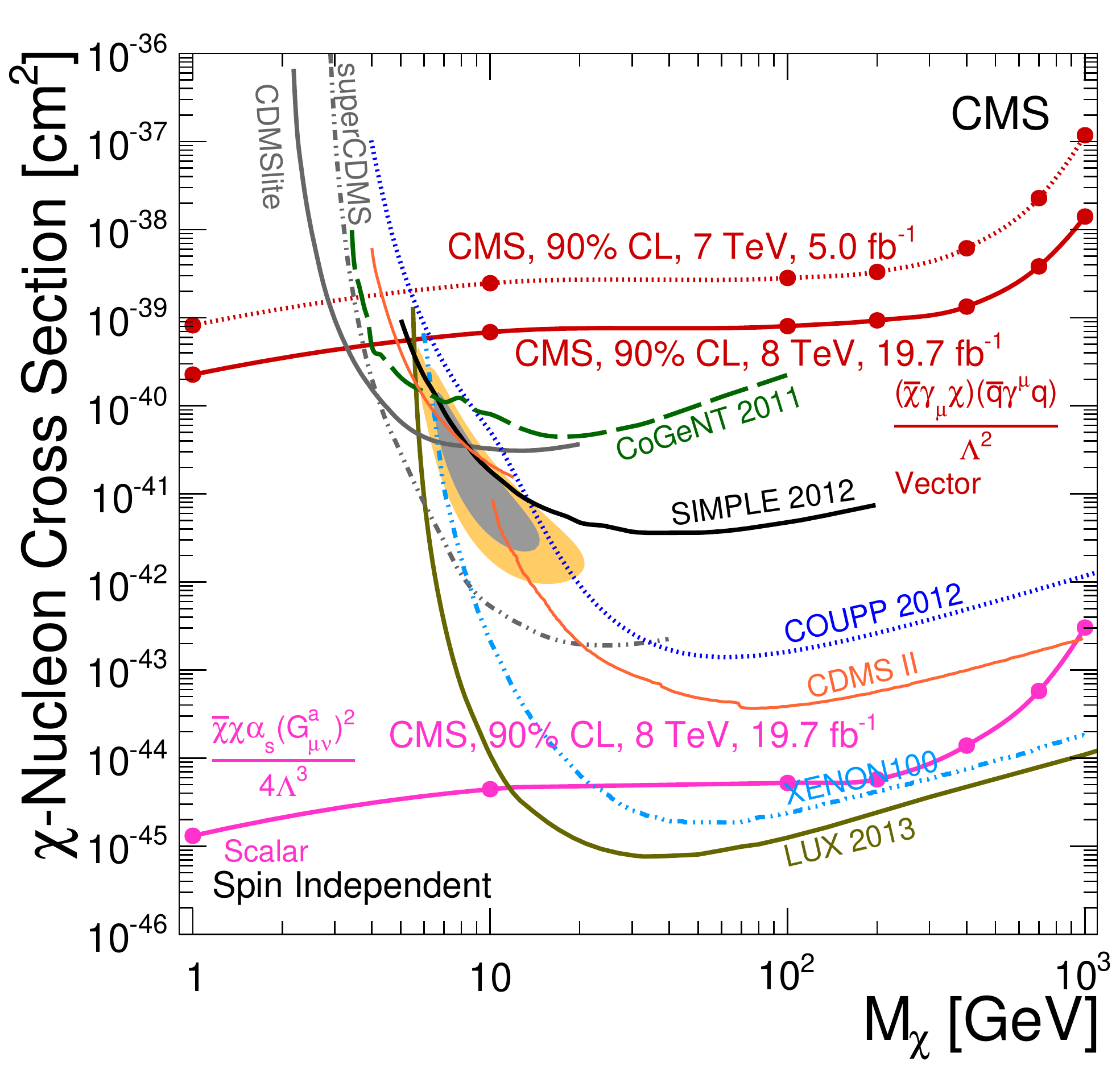}
  \includegraphics[width=\cmsFigWidth]{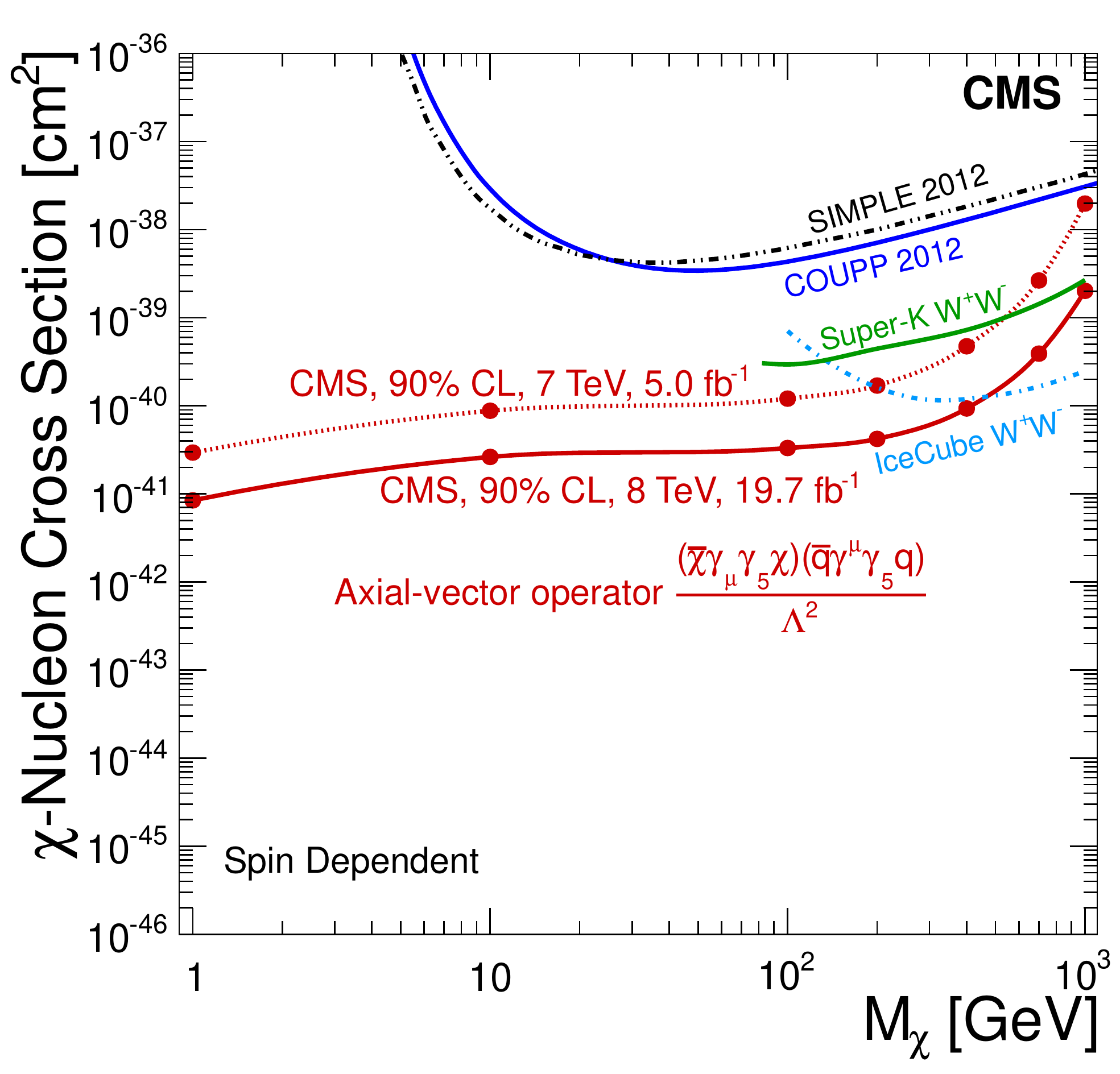}
  \caption{Upper limits on the DM-nucleon cross section, at
90\% CL, plotted against DM particle mass and compared with previously
published results.  \cmsLeft: limits for the
vector and scalar operators from the previous CMS analysis~\cite{bib:CMSEXO11059}, together with results from the
CoGeNT~\cite{bib:COGENT}, SIMPLE~\cite{SIMPLE2012},
COUPP~\cite{bib:COUPP2012}, CDMS~\cite{bib:CDMSII2010,bib:CDMSII2011}, SuperCDMS~\cite{SuperCDMS}, XENON100~\cite{bib:XENON100}, and LUX~\cite{bib:LUX} collaborations. The solid and hatched yellow contours show the 68\% and 90\% CL contours respectively for a possible signal from CDMS~\cite{bib:CDMSSi}.
\cmsRight: limits for the axial-vector operator from the previous CMS analysis~\cite{bib:CMSEXO11059},
together with results from the SIMPLE~\cite{SIMPLE2012},
COUPP~\cite{bib:COUPP2012}, Super-K~\cite{SUPERK}, and
IceCube~\cite{IceCube:2011aj} collaborations.
  \label{fig:DM_limits}}

\end{figure}

\begin{figure}[!Hhtb]
  \centering
  \includegraphics[width=\cmsFigWidth]{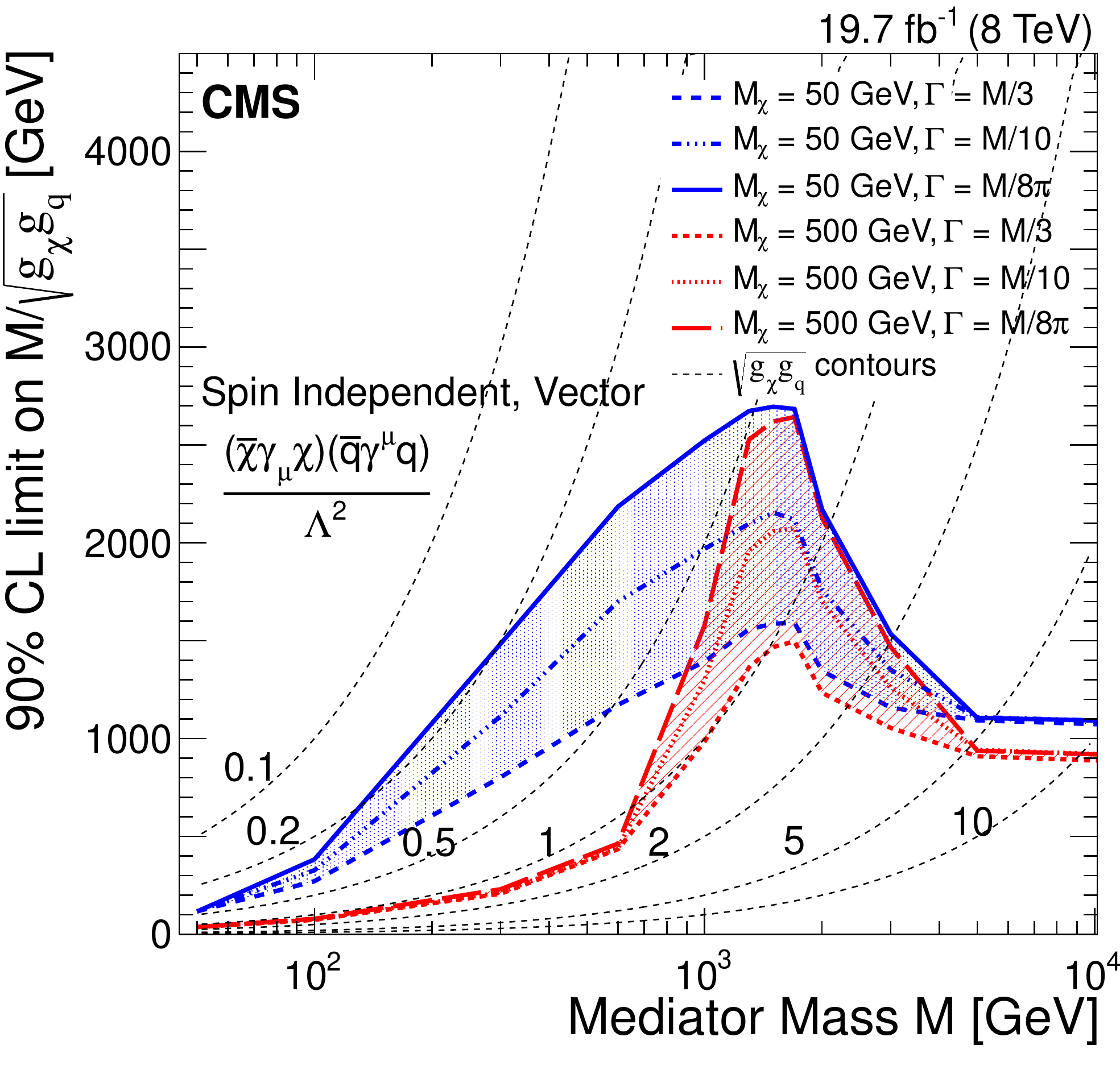}
  \caption{Observed limits on the mediator mass divided by coupling,
$M/\!\sqrt{g_{\chi}g_{\Pq}}$,
as a function of the mass of the
mediator, $M$, assuming vector interactions and a dark matter mass of
50\GeV (blue, filled) and 500\GeV (red, hatched). The width, $\Gamma$, of the
mediator is varied between $M/3$ and $M/8\pi$. The dashed lines show
contours of constant coupling $\sqrt{g_{\chi}g_{\Pq}}$.}
  \label{fig:DM_limits_scan}

\end{figure}

Lower limits on \MD in the ADD model, for different values of $\delta$, have
been obtained using LO cross section calculations,
and the application of NLO QCD
corrections, using $K$-factors, $K = \sigma_\mathrm{NLO}/\sigma_\mathrm{LO}$ of 1.4 for $\delta= \{2, 3\},$ 1.3 for $\delta = \{4,$ 5\}, and 1.2 for $\delta = 6$~\cite{Karg:2009xk}. Figure~\ref{fig:ADD_summary} shows
95\% CL limits at LO, compared to published results from ATLAS,
LEP, and the Tevatron. The ATLAS limits were produced using the full kinematic phase space, without any truncation applied to restrict the phase space to the region where the effective field theory is valid. The CMS limits are obtained using the truncated phase space, after discarding events for which the parton center of mass energy $\hat{s} > \MD^2$. The maximum difference in the cross section evaluated with and without the truncation was found to be 11\%. Table~\ref{tab:ADD_limits_MD} shows the expected and observed limits at LO and NLO for the ADD model.

\begin{figure}[!Hhtb]
  \centering
  \includegraphics[width=\cmsFigWidth] {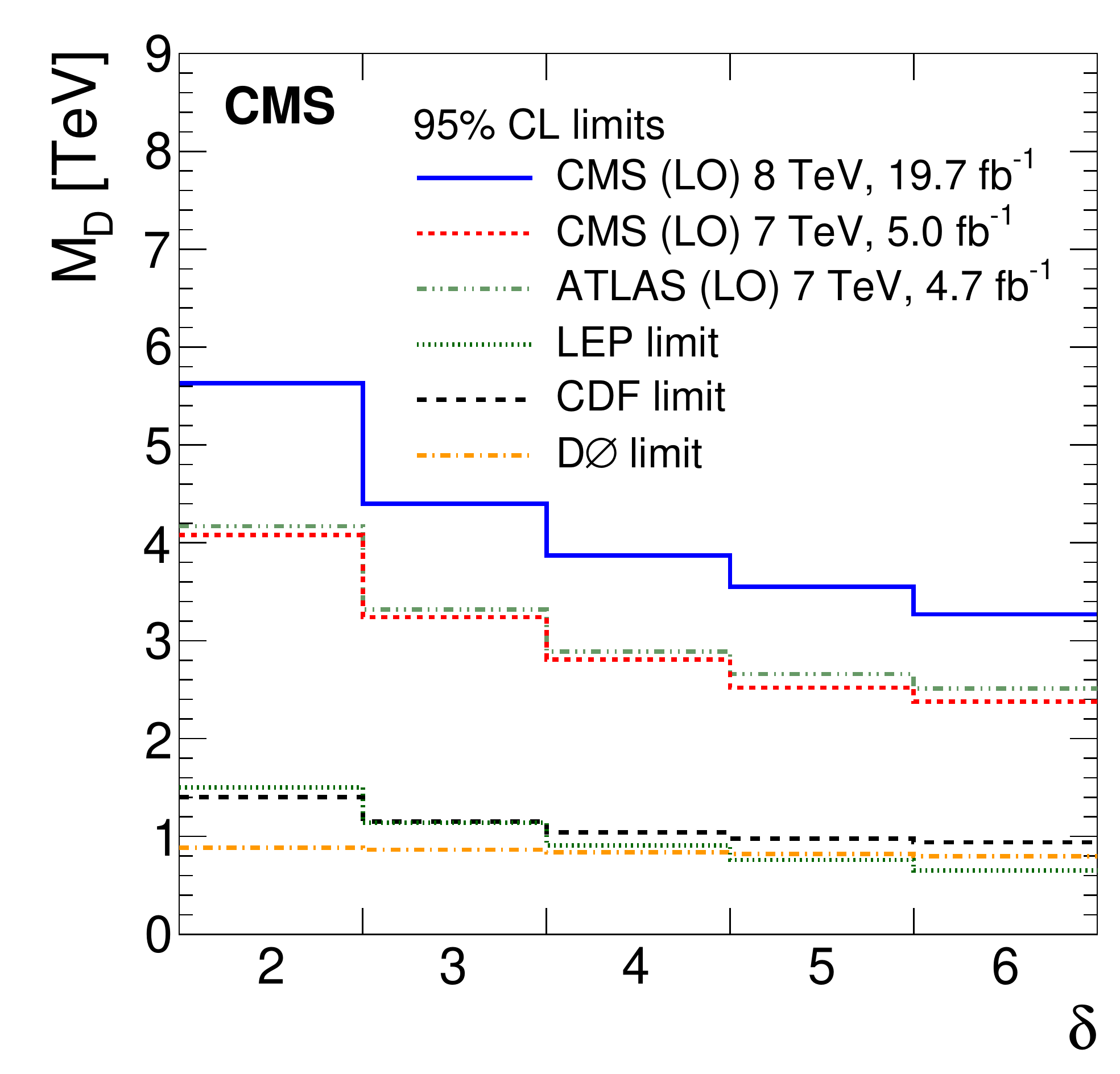}
   \caption{Lower limits at 95\% CL on \MD plotted against the number of extra
dimensions $\delta$, with results from the ATLAS~\cite{bib:ATLASMonoJet}, CMS~\cite{bib:CMSEXO11059}, LEP~\cite{bib:ALEPH,bib:OPAL,bib:DELPHI,bib:L3},
CDF~\cite{bib:CDFMonoPhoton}, and D\O~\cite{bib:D0MonoPhoton} collaborations.
         \label{fig:ADD_summary}}

\end{figure}

\begin{table*}[!Hhtb]  
        \centering
        \topcaption{Expected and observed 95\% CL lower limits on ADD model parameter $M_\mathrm{D}$ in \TeVns{} as a function of $\delta$ at LO and NLO.\label{tab:STAT_ADD_limits_NLO}}
        \begin{tabular}{c|cccc} \hline
          &\multicolumn{4}{c}{ LO limit on $M_\mathrm{D}$ (\TeVns{}) }  \\
          $\delta$& Expected limit & ${+}1\sigma$ & ${-}1\sigma$ &  Observed limit \\ \hline
             2 &   5.09 & 4.80 & 5.60 & 5.61 \\
             3 &   3.99 & 3.87 & 4.36 & 4.38 \\
             4 &   3.74 & 3.56 & 3.86 & 3.86 \\
             5 &   3.32 & 2.99 & 3.54 & 3.55 \\
             6 &   2.99 & 2.98 & 3.25 & 3.26 \\ \hline
          &\multicolumn{4}{c}{ NLO limit on \MD (\TeVns)}  \\
          $\delta$& Expected limit & ${+}1\sigma$ & ${-}1\sigma$ &  Observed limit \\ \hline
             2 &   5.53 & 5.21 & 6.08 & 6.09 \\
             3 &   4.34 & 4.21 & 4.74 & 4.77 \\
             4 &   3.85 & 3.66 & 3.97 & 3.97 \\
             5 &   3.49 & 3.14 & 3.72 & 3.73 \\
             6 &   3.24 & 3.23 & 3.52 & 3.53 \\ \hline
        \end{tabular}
	    \label{tab:ADD_limits_MD}

\end{table*}
Figure~\ref{fig:UP_limits} shows the expected and observed 95\% CL limits on
the cross-sections for scalar unparticles ($\mathrm{S} = 0$) with $d_\mathrm{U} = 1.5,$ 1.6, 1.7, 1.8, and
1.9 as a function of $\LambU$ for a fixed coupling constant $\lambda = 1$.
The observed 95\% CL limit $\LambU$ for these values of $d_\mathrm{U}$ is shown in
Table~\ref{tab:UP_limits_lambda}.

\begin{figure}[!Hh]
  \centering
  \includegraphics[width=\cmsFigWidth] {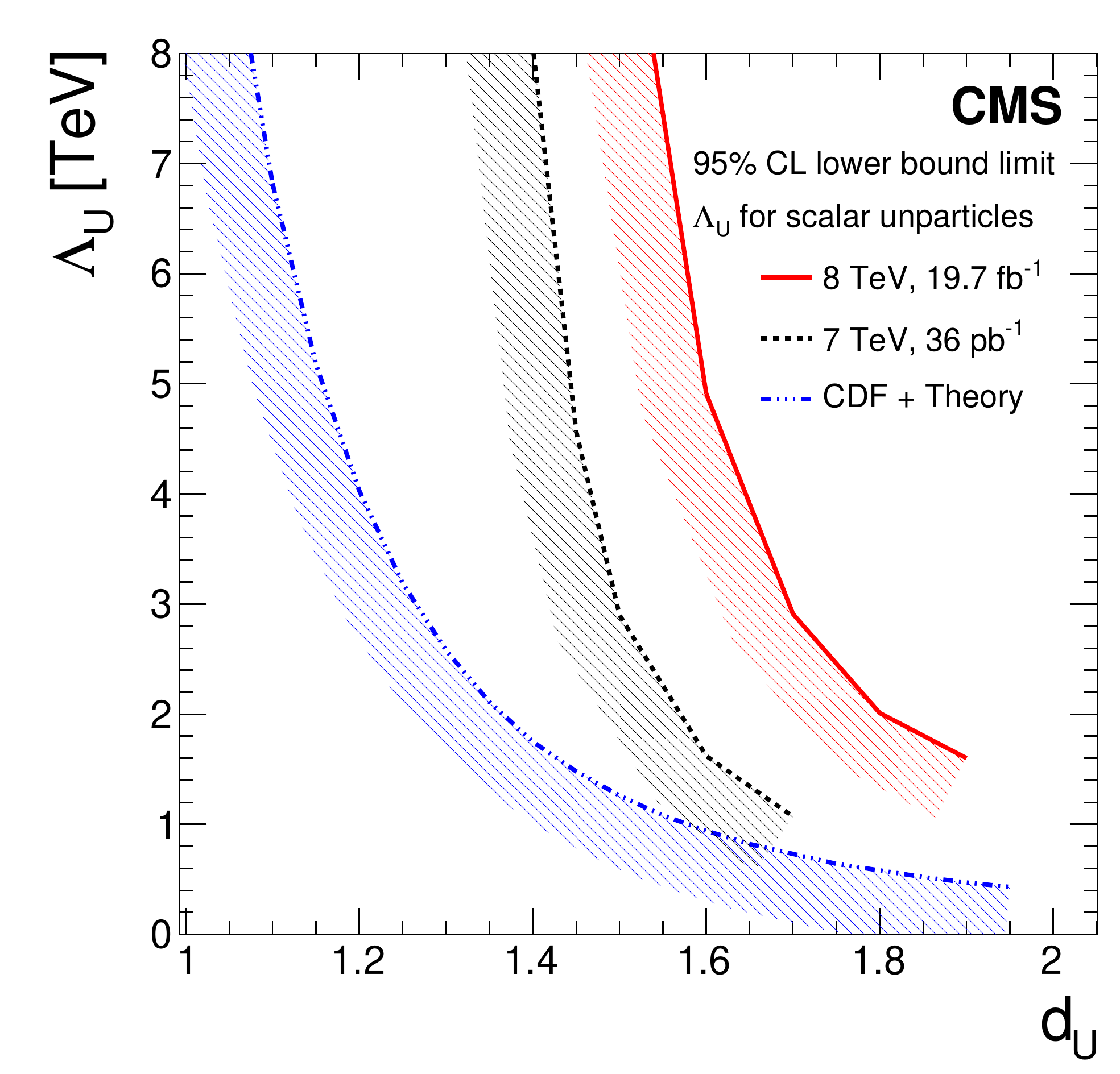}
   \caption{The expected and observed lower limits on the unparticle model parameters $\LambU$ as a function of $d_\mathrm{U}$ at 95\% CL, compared to previous results~\cite{bib:Kathrein,bib:CMS_EXO11003}. The shaded region indicates the side of the curve that is excluded.
         \label{fig:UP_limits}}

\end{figure}

\begin{table*}[!Hh]  
\centering
\topcaption{Expected and observed 95\% CL lower limits on $\LambU$ (in \TeVns)
for scalar unparticles with $d_\mathrm{U} = $1.5, 1.6, 1.7, 1.8 and 1.9 and a fixed
coupling constant $\lambda = 1$.}
\label{tab:UP_limits_lambda}
 \begin{tabular}{ccccc} \hline
 $d_\mathrm{U}$  & Expected limit on $\LambU$ (TeVns{}) & ${+}1\sigma$ & ${-}1\sigma$ & Observed limit on $\LambU$ (\TeVns)\\ \hline
1.5 &  7.88 & 6.63 & 8.39 & 10.00 \\
1.6 &  3.89 & 2.51 & 4.88 & 4.91 \\
1.7 &  2.63 & 2.09 & 2.89 & 2.91 \\
1.8 &  1.91 & 1.76 & 1.98 & 2.01 \\
1.9 &  1.41 & 0.88 & 1.46 & 1.60 \\ \hline

\end{tabular}

\end{table*}

\section{Summary}

A search for particle dark matter, large extra dimensions, and
unparticle production has been performed in the monojet
channel using a data sample of proton-proton collisions at $\sqrt{s} = 8$
\TeV corresponding to an integrated luminosity of 19.7\fbinv.
The dominant
backgrounds to this topology are from $\cPZ(\cPgn\cPgn)$+jets and $\wellnubr+$jets events, and are
estimated from data samples of \Zmumu and \Wmunu events, respectively. The data
are found to be in agreement with expected contributions from standard model
processes. Limits are set on the DM-nucleon scattering cross
section assuming vector, axial-vector,
and scalar operators. Limits are also set on the fundamental Planck scale \MD in the ADD model of large extra dimensions and on the unparticle model parameter $\LambU$. Compared to previous CMS publications in this channel, the lower limits on \MD represent an approximately 40\% improvement, and the lower limits on the unparticle model parameter $\LambU$ represent an improvement by a factor of roughly 3. The upper limit on the DM-nucleon cross section has been reduced from $8.79\times 10^{-41}\cm^{2}$ to $2.70\times 10^{-41}\cm^{2}$ for the axial-vector operator and from $2.47\times 10^{-39}\cm^{2}$ to $7.06\times 10^{-40}\cm^{2}$ for the vector operator for a particle DM mass of 10\GeV.
The constraints on ADD models and unparticles are the most stringent limits in this channel and those on the DM-nucleon scattering cross section are an improvement over previous collider results.

\begin{acknowledgments}
We congratulate our colleagues in the CERN accelerator departments for the excellent performance of the LHC and thank the technical and administrative staffs at CERN and at other CMS institutes for their contributions to the success of the CMS effort. In addition, we gratefully acknowledge the computing centres and personnel of the Worldwide LHC Computing Grid for delivering so effectively the computing infrastructure essential to our analyses. Finally, we acknowledge the enduring support for the construction and operation of the LHC and the CMS detector provided by the following funding agencies: BMWFW and FWF (Austria); FNRS and FWO (Belgium); CNPq, CAPES, FAPERJ, and FAPESP (Brazil); MES (Bulgaria); CERN; CAS, MoST, and NSFC (China); COLCIENCIAS (Colombia); MSES and CSF (Croatia); RPF (Cyprus); MoER, ERC IUT and ERDF (Estonia); Academy of Finland, MEC, and HIP (Finland); CEA and CNRS/IN2P3 (France); BMBF, DFG, and HGF (Germany); GSRT (Greece); OTKA and NIH (Hungary); DAE and DST (India); IPM (Iran); SFI (Ireland); INFN (Italy); NRF and WCU (Republic of Korea); LAS (Lithuania); MOE and UM (Malaysia); CINVESTAV, CONACYT, SEP, and UASLP-FAI (Mexico); MBIE (New Zealand); PAEC (Pakistan); MSHE and NSC (Poland); FCT (Portugal); JINR (Dubna); MON, RosAtom, RAS and RFBR (Russia); MESTD (Serbia); SEIDI and CPAN (Spain); Swiss Funding Agencies (Switzerland); MST (Taipei); ThEPCenter, IPST, STAR and NSTDA (Thailand); TUBITAK and TAEK (Turkey); NASU and SFFR (Ukraine); STFC (United Kingdom); DOE and NSF (USA).

Individuals have received support from the Marie-Curie programme and the European Research Council and EPLANET (European Union); the Leventis Foundation; the A. P. Sloan Foundation; the Alexander von Humboldt Foundation; the Belgian Federal Science Policy Office; the Fonds pour la Formation \`a la Recherche dans l'Industrie et dans l'Agriculture (FRIA-Belgium); the Agentschap voor Innovatie door Wetenschap en Technologie (IWT-Belgium); the Ministry of Education, Youth and Sports (MEYS) of the Czech Republic; the Council of Science and Industrial Research, India; the HOMING PLUS programme of Foundation for Polish Science, cofinanced from European Union, Regional Development Fund; the Compagnia di San Paolo (Torino); the Consorzio per la Fisica (Trieste); MIUR project 20108T4XTM (Italy); the Thalis and Aristeia programmes cofinanced by EU-ESF and the Greek NSRF; and the National Priorities Research Program by Qatar National Research Fund.
\end{acknowledgments}
\bibliography{auto_generated}

\providecommand{\href}[2]{#2}\begingroup\raggedright\begin{thebibliography}{10}%
\makeatletter
\providecommand{\hrefCMSnoop }[0]{\@secondoftwo}%
\makeatother
\providecommand{\doi}{\texttt{doi:}\begingroup \urlstyle{tt}\Url}

\bibitem{DarkMatterReview}
\hrefCMSnoop {}{V.~Trimble, ``Existence and Nature of Dark Matter in the
  Universe'',} \textit{ Ann. Rev. Astron. Astrophys.} \textbf{ 25} (1987) 425,
  \href{http://dx.doi.org/10.1146/annurev.aa.25.090187.002233}{\doi{10.1146/annurev.aa.25.090187.002233}}.

\bibitem{DMGeneral}
\hrefCMSnoop {}{J.~L. Feng, ``{Dark Matter Candidates from Particle Physics and
  Methods of Detection}'',} \textit{ Ann. Rev. Astron. Astrophys.} \textbf{ 48}
  (2010) 495,
  \href{http://dx.doi.org/10.1146/annurev-astro-082708-101659}{\doi{10.1146/annurev-astro-082708-101659}},
\href{http://www.arXiv.org/abs/1003.0904}{\texttt{arXiv:1003.0904}}.

\bibitem{bib:WMAP9}
\hrefCMSnoop {}{G.~Hinshaw {et~al.}, ``Nine-year Wilkinson Microwave Anisotropy
  Probe (WMAP) observations: Cosmological parameter results'',} \textit{ The
  Astrophysical Journal Supplement Series} \textbf{ 208} (2013), no.~2, 19,
  \href{http://dx.doi.org/10.1088/0067-0049/208/2/19}{\doi{10.1088/0067-0049/208/2/19}}.

\bibitem{bib:Planck}
\hrefCMSnoop {}{{Planck} Collaboration, ``{Planck 2013 results. XVI.
  Cosmological parameters}'',} \textit{ Astron. Astrophys.} \textbf{ 571}
  (2014) A16,
  \href{http://dx.doi.org/10.1051/0004-6361/201321591}{\doi{10.1051/0004-6361/201321591}},
\href{http://www.arXiv.org/abs/1303.5076}{\texttt{arXiv:1303.5076}}.

\bibitem{bib:SUSY}
\hrefCMSnoop {}{G.~R. Farrar and P.~Fayet, ``Phenomenology of the production,
  decay, and detection of new hadronic states associated with supersymmetry'',}
  \textit{ Phys. Lett. B} \textbf{ 76} (1978) 575,
\href{http://dx.doi.org/10.1016/0370-2693(78)90858-4}{\doi{10.1016/0370-2693(78)90858-4}}.

\bibitem{bib:TMTait}
M.~Beltran\hrefCMSnoop {}{ {et~al.}, ``Maverick dark matter at colliders'',}
  \textit{ JHEP} \textbf{ 09} (2010) 037,
  \href{http://dx.doi.org/10.1007/JHEP09(2010)037}{\doi{10.1007/JHEP09(2010)037}},
\href{http://www.arXiv.org/abs/1002.4137}{\texttt{arXiv:1002.4137}}.

\bibitem{bib:TMTait2}
J.~Goodman\hrefCMSnoop {}{ {et~al.}, ``Constraints on dark matter from
  colliders'',} \textit{ Phys. Rev. D} \textbf{ 82} (2010) 116010,
  \href{http://dx.doi.org/10.1103/PhysRevD.82.116010}{\doi{10.1103/PhysRevD.82.116010}},
\href{http://www.arXiv.org/abs/1008.1783}{\texttt{arXiv:1008.1783}}.

\bibitem{TimTaitB}
J.~Goodman\hrefCMSnoop {}{ {et~al.}, ``{Constraints on light Majorana dark
  matter from colliders}'',} \textit{ Phys. Lett. B} \textbf{ 695} (2011) 185,
  \href{http://dx.doi.org/10.1016/j.physletb.2010.11.009}{\doi{10.1016/j.physletb.2010.11.009}},
\href{http://www.arXiv.org/abs/1005.1286}{\texttt{arXiv:1005.1286}}.

\bibitem{bib:RoniHarnik}
\hrefCMSnoop {}{Y.~Bai, P.~J. Fox, and R.~Harnik, ``{The Tevatron at the
  frontier of dark matter direct detection}'',} \textit{ JHEP} \textbf{ 12}
  (2010) 048,
  \href{http://dx.doi.org/10.1007/JHEP12(2010)048}{\doi{10.1007/JHEP12(2010)048}},
\href{http://www.arXiv.org/abs/1005.3797}{\texttt{arXiv:1005.3797}}.

\bibitem{bib:CDFmonojet}
\hrefCMSnoop {}{{CDF} Collaboration, ``A Search for Dark Matter in Events with
  One Jet and Missing Transverse Energy in $ p\bar{p}$ Collisions at $\sqrt{s}
  = 1.96$ {TeV}'',} \textit{ Phys. Rev. Lett.} \textbf{ 108} (2012) 211804,
  \href{http://dx.doi.org/10.1103/PhysRevLett.108.211804}{\doi{10.1103/PhysRevLett.108.211804}},
\href{http://www.arXiv.org/abs/1203.0742}{\texttt{arXiv:1203.0742}}.

\bibitem{bib:CMSEXO11059}
\hrefCMSnoop {}{{CMS} Collaboration, ``{Search for dark matter and large extra
  dimensions in monojet events in pp collisions at $\sqrt{s}=7$ TeV}'',}
  \textit{ JHEP} \textbf{ 09} (2012) 094,
  \href{http://dx.doi.org/10.1007/JHEP09(2012)094}{\doi{10.1007/JHEP09(2012)094}},
\href{http://www.arXiv.org/abs/1206.5663}{\texttt{arXiv:1206.5663}}.

\bibitem{bib:ATLAS2012ky}
\hrefCMSnoop {}{{ATLAS} Collaboration, ``{Search for dark matter candidates and
  large extra dimensions in events with a jet and missing transverse momentum
  with the ATLAS detector}'',} \textit{ JHEP} \textbf{ 04} (2013) 075,
  \href{http://dx.doi.org/10.1007/JHEP04(2013)075}{\doi{10.1007/JHEP04(2013)075}},
\href{http://www.arXiv.org/abs/1210.4491}{\texttt{arXiv:1210.4491}}.

\bibitem{bib:ADD1}
\hrefCMSnoop {}{N.~Arkani-Hamed, S.~Dimopoulos, and G.~Dvali, ``The hierarchy
  problem and new dimensions at a millimeter'',} \textit{ Phys. Lett. B}
  \textbf{ 429} (1998) 263,
  \href{http://dx.doi.org/10.1016/S0370-2693(98)00466-3}{\doi{10.1016/S0370-2693(98)00466-3}},
  \href{http://www.arXiv.org/abs/hep-ph/9803315}{\texttt{arXiv:hep-ph/9803315}}.

\bibitem{ADDPRD}
\hrefCMSnoop {}{N.~Arkani-Hamed, S.~Dimopoulos, and G.~Dvali, ``{Phenomenology,
  astrophysics and cosmology of theories with submillimeter dimensions and TeV
  scale quantum gravity}'',} \textit{ Phys. Rev. D} \textbf{ 59} (1999) 086004,
  \href{http://dx.doi.org/10.1103/PhysRevD.59.086004}{\doi{10.1103/PhysRevD.59.086004}},
\href{http://www.arXiv.org/abs/hep-ph/9807344}{\texttt{arXiv:hep-ph/9807344}}.

\bibitem{Antoniadis}
\hrefCMSnoop {}{I.~Antoniadis, K.~Benakli, and M.~Quiros, ``{Direct collider
  signatures of large extra dimensions}'',} \textit{ Phys. Lett. B} \textbf{
  460} (1999) 176,
  \href{http://dx.doi.org/10.1016/S0370-2693(99)00764-9}{\doi{10.1016/S0370-2693(99)00764-9}},
\href{http://www.arXiv.org/abs/hep-ph/9905311}{\texttt{arXiv:hep-ph/9905311}}.

\bibitem{ADDGiudice}
\hrefCMSnoop {}{G.~Giudice, R.~Rattazzi, and J.~Wells, ``Quantum gravity and
  extra dimensions at high-energy colliders'',} \textit{ Nucl. Phys. B}
  \textbf{ 544} (1999) 3,
  \href{http://dx.doi.org/10.1016/S0550-3213(99)00044-9}{\doi{10.1016/S0550-3213(99)00044-9}},
  \href{http://www.arXiv.org/abs/hep-ph/9811291}{\texttt{arXiv:hep-ph/9811291}}.

\bibitem{ADDPeskin}
\hrefCMSnoop {}{E.~Mirabelli, M.~Perelstein, and M.~Peskin, ``Collider
  signatures of new large space dimensions'',} \textit{ Phys. Rev. Lett.}
  \textbf{ 82} (1999) 2236,
  \href{http://dx.doi.org/10.1103/PhysRevLett.82.2236}{\doi{10.1103/PhysRevLett.82.2236}},
  \href{http://www.arXiv.org/abs/hep-ph/9811337}{\texttt{arXiv:hep-ph/9811337}}.

\bibitem{Witten1981267}
\hrefCMSnoop {}{E.~Witten, ``Mass hierarchies in supersymmetric theories'',}
  \textit{ Phys. Lett. B} \textbf{ 105} (1981) 267,
  \href{http://dx.doi.org/10.1016/0370-2693(81)90885-6}{\doi{10.1016/0370-2693(81)90885-6}}.

\bibitem{bib:OPAL}
\hrefCMSnoop {}{{OPAL} Collaboration, ``Photonic events with missing energy in
  e$^{+}$e$^{-}$ collisions at $\sqrt{s} = 189$ {GeV}'',} \textit{ Eur. Phys.
  J. C} \textbf{ 18} (2000) 253,
  \href{http://dx.doi.org/10.1007/s100520000522}{\doi{10.1007/s100520000522}},
\href{http://www.arXiv.org/abs/hep-ex/0005002}{\texttt{arXiv:hep-ex/0005002}}.

\bibitem{bib:ALEPH}
\hrefCMSnoop {}{{ALEPH} Collaboration, ``Single- and multi-photon production in
  e$^{+}$e$^{-}$ collisions at $\sqrt{s}$ up to 209 {GeV}'',} \textit{ Eur.
  Phys. J. C} \textbf{ 28} (2003) 1,
  \href{http://dx.doi.org/10.1140/epjc/s2002-01129-7}{\doi{10.1140/epjc/s2002-01129-7}}.

\bibitem{bib:L3}
\hrefCMSnoop {}{{L3} Collaboration, ``Single- and multi-photon events with
  missing energy in e$^{+}$e$^{-}$ collisions at {LEP}'',} \textit{ Phys. Lett.
  B} \textbf{ 587} (2004) 16,
  \href{http://dx.doi.org/10.1016/j.physletb.2004.01.010}{\doi{10.1016/j.physletb.2004.01.010}},
\href{http://www.arXiv.org/abs/hep-ex/0402002}{\texttt{arXiv:hep-ex/0402002}}.

\bibitem{bib:CDFMonoPhoton}
\hrefCMSnoop {}{{CDF} Collaboration, ``Search for Large Extra Dimensions in
  Final States Containing One Photon or Jet and Large Missing Transverse Energy
  Produced in \ppbar Collisions at $\sqrt{s} = 1.96$ {TeV}'',} \textit{ Phys.
  Rev. Lett.} \textbf{ 101} (2008) 181602,
  \href{http://dx.doi.org/10.1103/PhysRevLett.101.181602}{\doi{10.1103/PhysRevLett.101.181602}},
  \href{http://www.arXiv.org/abs/0807.3132}{\texttt{arXiv:0807.3132}}.

\bibitem{bib:D0MonoPhoton}
\hrefCMSnoop {}{{D0} Collaboration, ``Search for Large Extra Dimensions via
  Single Photon plus Missing Energy Final States at $\sqrt{s} = 1.96$ {TeV}'',}
  \textit{ Phys. Rev. Lett.} \textbf{ 101} (2008) 011601,
  \href{http://dx.doi.org/10.1103/PhysRevLett.101.011601}{\doi{10.1103/PhysRevLett.101.011601}},
  \href{http://www.arXiv.org/abs/0803.2137}{\texttt{arXiv:0803.2137}}.

\bibitem{bib:CMS_EXO11003}
\hrefCMSnoop {}{{CMS} Collaboration, ``{Search for New Physics with a Mono-Jet
  and Missing Transverse Energy in pp Collisions at $\sqrt{s} = 7$ TeV}'',}
  \textit{ Phys. Rev. Lett.} \textbf{ 107} (2011) 201804,
  \href{http://dx.doi.org/10.1103/PhysRevLett.107.201804}{\doi{10.1103/PhysRevLett.107.201804}},
  \href{http://www.arXiv.org/abs/1106.4775}{\texttt{arXiv:1106.4775}}.

\bibitem{bib:ATLASMonoJet}
\hrefCMSnoop {}{{ATLAS} Collaboration, ``{Search for new phenomena with the
  monojet and missing transverse momentum signature using the ATLAS detector in
  $\sqrt{s} = 7$ TeV proton-proton collisions}'',} \textit{ Phys. Lett. B}
  \textbf{ 705} (2011) 294,
  \href{http://dx.doi.org/10.1016/j.physletb.2011.10.006}{\doi{10.1016/j.physletb.2011.10.006}},
\href{http://www.arXiv.org/abs/1106.5327}{\texttt{arXiv:1106.5327}}.

\bibitem{bib:Unp}
\hrefCMSnoop {}{H.~Georgi, ``Unparticle Physics'',} \textit{ Phys. Rev. Lett.}
  \textbf{ 98} (2007) 221601,
  \href{http://dx.doi.org/10.1103/PhysRevLett.98.221601}{\doi{10.1103/PhysRevLett.98.221601}},
\href{http://www.arXiv.org/abs/hep-ph/0703260}{\texttt{arXiv:hep-ph/0703260}}.

\bibitem{bib:CMS_TDR}
\hrefCMSnoop {}{{CMS} Collaboration, ``The CMS experiment at the CERN LHC'',}
  \textit{ JINST} \textbf{ 3} (2008) S08004,
  \href{http://dx.doi.org/10.1088/1748-0221/3/08/S08004}{\doi{10.1088/1748-0221/3/08/S08004}}.

\bibitem{bib:ANA_PF}
\href {http://cdsweb.cern.ch/record/1194487}{{CMS} Collaboration,
  ``Particle--Flow Event Reconstruction in {CMS} and Performance for Jets,
  Taus, and {\MET}'',} CMS Physics Analysis Summary CMS-PAS-PFT-09-001, 2009.

\bibitem{CMS-PAS-PFT-10-001}
\href {http://cdsweb.cern.ch/record/1247373}{{CMS} Collaboration,
  ``Commissioning of the Particle-flow Event Reconstruction with the first
  {LHC} collisions recorded in the {CMS} detector'',} CMS Physics Analysis
  Summary CMS-PAS-PFT-10-001, 2010.

\bibitem{bib:ANA_AK}
\hrefCMSnoop {}{M.~Cacciari, G.~P. Salam, and G.~Soyez, ``The anti-$k_t$ jet
  clustering algorithm'',} \textit{ JHEP} \textbf{ 04} (2008) 063,
  \href{http://dx.doi.org/10.1088/1126-6708/2008/04/063}{\doi{10.1088/1126-6708/2008/04/063}},
  \href{http://www.arXiv.org/abs/0802.1189}{\texttt{arXiv:0802.1189}}.

\bibitem{Jetresolution}
\hrefCMSnoop {}{{CMS} Collaboration, ``{Energy calibration and resolution of
  the CMS electromagnetic calorimeter in pp collisions at $\sqrt{s}$ = 7
  TeV}'',} \textit{ JINST} \textbf{ 8} (2013) P09009,
  \href{http://dx.doi.org/10.1088/1748-0221/8/09/P09009}{\doi{10.1088/1748-0221/8/09/P09009}},
\href{http://www.arXiv.org/abs/1306.2016}{\texttt{arXiv:1306.2016}}.

\bibitem{JETJINST}
\hrefCMSnoop {}{{CMS} Collaboration, ``{Determination of jet energy calibration
  and transverse momentum resolution in CMS}'',} \textit{ JINST} \textbf{ 6}
  (2011) P11002,
  \href{http://dx.doi.org/10.1088/1748-0221/6/11/P11002}{\doi{10.1088/1748-0221/6/11/P11002}},
\href{http://www.arXiv.org/abs/1107.4277}{\texttt{arXiv:1107.4277}}.

\bibitem{Fastjet}
\hrefCMSnoop {}{M.~Cacciari, G.~P. Salam, and G.~Soyez, ``{FastJet} user
  manual'',} \textit{ Eur. Phys. J. C} \textbf{ 72} (2012) 1896,
  \href{http://dx.doi.org/10.1140/epjc/s10052-012-1896-2}{\doi{10.1140/epjc/s10052-012-1896-2}},
\href{http://www.arXiv.org/abs/1111.6097}{\texttt{arXiv:1111.6097}}.

\bibitem{bib:ANA_Tk}
\hrefCMSnoop {}{{CMS} Collaboration, ``{CMS} tracking performance results from
  early {LHC} operation'',} \textit{ Eur. Phys. J. C} \textbf{ 70} (2010) 1165,
  \href{http://dx.doi.org/10.1140/epjc/s10052-010-1491-3}{\doi{10.1140/epjc/s10052-010-1491-3}},
  \href{http://www.arXiv.org/abs/1007.1988}{\texttt{arXiv:1007.1988}}.

\bibitem{bib:muons}
\hrefCMSnoop {}{{CMS} Collaboration, ``Performance of {CMS} muon reconstruction
  in pp collision events at {$\sqrt{s} = 7$\TeV}'',} \textit{ J. Instrum.}
  \textbf{ 7} (2012) P10002,
  \href{http://dx.doi.org/10.1088/1748-0221/7/10/P10002}{\doi{10.1088/1748-0221/7/10/P10002}}.

\bibitem{bib:HPStaus}
\hrefCMSnoop {}{{CMS} Collaboration, ``Performance of $\tau$-lepton
  reconstruction and identification in {CMS}'',} \textit{ J. Instrum.} \textbf{
  7} (2012) P01001,
  \href{http://dx.doi.org/10.1088/1748-0221/7/01/P01001}{\doi{10.1088/1748-0221/7/01/P01001}}.

\bibitem{bib:GEN_Mg}
J.~Alwall\hrefCMSnoop {}{ {et~al.}, ``MadGraph/MadEvent v4: the new web
  generation'',} \textit{ JHEP} \textbf{ 09} (2007) 028,
  \href{http://dx.doi.org/10.1088/1126-6708/2007/09/028}{\doi{10.1088/1126-6708/2007/09/028}},
  \href{http://www.arXiv.org/abs/0706.2334}{\texttt{arXiv:0706.2334}}.

\bibitem{bib:GEN_Py6}
\hrefCMSnoop {}{T.~Sj{\"o}strand, S.~Mrenna, and P.~Z. Skands, ``PYTHIA 6.4
  physics and manual'',} \textit{ JHEP} \textbf{ 05} (2006) 026,
  \href{http://dx.doi.org/10.1088/1126-6708/2006/05/026}{\doi{10.1088/1126-6708/2006/05/026}},
  \href{http://www.arXiv.org/abs/hep-ph/0603175}{\texttt{arXiv:hep-ph/0603175}}.

\bibitem{bib:Z2star}
\hrefCMSnoop {}{R.~Field, ``{Early LHC Underlying Event Data - Findings and
  Surprises}'',} (2010).
\href{http://www.arXiv.org/abs/1010.3558}{\texttt{arXiv:1010.3558}}.

\bibitem{bib:SYST_CTEQ6M}
J.~Pumplin\hrefCMSnoop {}{ {et~al.}, ``New generation of parton distributions
  with uncertainties from global QCD analysis'',} \textit{ JHEP} \textbf{ 07}
  (2002) 012,
  \href{http://dx.doi.org/10.1088/1126-6708/2002/07/012}{\doi{10.1088/1126-6708/2002/07/012}},
  \href{http://www.arXiv.org/abs/hep-ph/0201195}{\texttt{arXiv:hep-ph/0201195}}.

\bibitem{bib:GEN_PY8}
\hrefCMSnoop {}{T.~Sj{\"o}strand, S.~Mrenna, and P.~Z. Skands, ``A brief
  introduction to PYTHIA 8.1'',} \textit{ Comput. Phys. Commun.} \textbf{ 178}
  (2008) 852,
  \href{http://dx.doi.org/10.1016/j.cpc.2008.01.036}{\doi{10.1016/j.cpc.2008.01.036}},
  \href{http://www.arXiv.org/abs/0710.3820}{\texttt{arXiv:0710.3820}}.

\bibitem{bib:GEN_Ask}
S.~Ask\hrefCMSnoop {}{ {et~al.}, ``Real emission and virtual exchange of
  gravitons and unparticles in PYTHIA8'',} \textit{ Comput. Phys. Commun.}
  \textbf{ 181} (2010) 1593,
  \href{http://dx.doi.org/10.1016/j.cpc.2010.05.013}{\doi{10.1016/j.cpc.2010.05.013}},
  \href{http://www.arXiv.org/abs/0912.4233}{\texttt{arXiv:0912.4233}}.

\bibitem{TuneFourC}
\hrefCMSnoop {}{R.~Corke and T.~Sj{\"o}strand, ``{Interleaved parton showers
  and tuning prospects}'',} \textit{ JHEP} \textbf{ 03} (2011) 032,
  \href{http://dx.doi.org/10.1007/JHEP03(2011)032}{\doi{10.1007/JHEP03(2011)032}},
\href{http://www.arXiv.org/abs/1011.1759}{\texttt{arXiv:1011.1759}}.

\bibitem{bib:MG5}
J.~Alwall\hrefCMSnoop {}{ {et~al.}, ``{MadGraph 5: going beyond}'',} \textit{
  JHEP} \textbf{ 06} (2011) 128,
  \href{http://dx.doi.org/10.1007/JHEP06(2011)128}{\doi{10.1007/JHEP06(2011)128}},
\href{http://www.arXiv.org/abs/1106.0522}{\texttt{arXiv:1106.0522}}.

\bibitem{Alwall:2014hca}
J.~Alwall\hrefCMSnoop {}{ {et~al.}, ``The automated computation of tree-level
  and next-to-leading order differential cross sections, and their matching to
  parton shower simulations'',} \textit{ JHEP} \textbf{ 07} (2014) 079,
  \href{http://dx.doi.org/10.1007/JHEP07(2014)079}{\doi{10.1007/JHEP07(2014)079}},
\href{http://www.arXiv.org/abs/1405.0301}{\texttt{arXiv:1405.0301}}.

\bibitem{bib:powheg}
\hrefCMSnoop {}{S.~Frixione, P.~Nason, and C.~Oleari, ``{Matching NLO QCD
  computations with parton shower simulations: the POWHEG method}'',} \textit{
  JHEP} \textbf{ 11} (2007) 070,
  \href{http://dx.doi.org/10.1088/1126-6708/2007/11/070}{\doi{10.1088/1126-6708/2007/11/070}},
\href{http://www.arXiv.org/abs/0709.2092}{\texttt{arXiv:0709.2092}}.

\bibitem{Alioli:2009je}
\hrefCMSnoop {}{S.~Alioli, P.~Nason, C.~Oleari, and E.~Re, ``{NLO single-top
  production matched with shower in POWHEG: $s$- and $t$-channel
  contributions}'',} \textit{ JHEP} \textbf{ 09} (2009) 111,
  \href{http://dx.doi.org/10.1088/1126-6708/2009/09/111}{\doi{10.1088/1126-6708/2009/09/111}},
  \href{http://www.arXiv.org/abs/0907.4076}{\texttt{arXiv:0907.4076}}.
[Erratum: \DOI{10.1007/JHEP02(2010)011}].

\bibitem{Agostinelli2003250}
\hrefCMSnoop {}{{GEANT4} Collaboration, ``{GEANT4}---a simulation toolkit'',}
  \textit{ Nucl. Instrum. Meth. A} \textbf{ 506} (2003) 250,
\href{http://dx.doi.org/10.1016/S0168-9002(03)01368-8}{\doi{10.1016/S0168-9002(03)01368-8}}.

\bibitem{bib:GEN_GEANT4}
\hrefCMSnoop {}{J.~Allison {et~al.}, ``{GEANT4 developments and
  applications}'',} \textit{ IEEE Trans. Nucl. Sci.} \textbf{ 53} (2006) 270,
\href{http://dx.doi.org/10.1109/TNS.2006.869826}{\doi{10.1109/TNS.2006.869826}}.

\bibitem{bib:PDF4LHC}
M.~Botje\hrefCMSnoop {}{ {et~al.}, ``{The PDF4LHC Working Group Interim
  Recommendations}'',} (2011).
\href{http://www.arXiv.org/abs/1101.0538}{\texttt{arXiv:1101.0538}}.

\bibitem{bib:PDF4LHC2}
\hrefCMSnoop {}{S.~Alekhin {et~al.}, ``{The PDF4LHC Working Group Interim
  Report}'',} (2011).
\href{http://www.arXiv.org/abs/1101.0536}{\texttt{arXiv:1101.0536}}.

\bibitem{bib:BKG_PDG}
\hrefCMSnoop {}{{Particle Data Group} Collaboration, ``Review of Particle
  Physics'',} \textit{ J. Phys. G} \textbf{ 37} (2010) 075021,
\href{http://dx.doi.org/10.1088/0954-3899/37/7A/075021}{\doi{10.1088/0954-3899/37/7A/075021}}.

\bibitem{bib:tagprobe}
\hrefCMSnoop {}{{CMS} Collaboration, ``{Measurements of Inclusive W and Z Cross
  Sections in pp Collisions at $\sqrt{s} = 7$ TeV}'',} \textit{ JHEP} \textbf{
  01} (2011) 080,
  \href{http://dx.doi.org/10.1007/JHEP01(2011)080}{\doi{10.1007/JHEP01(2011)080}},
\href{http://www.arXiv.org/abs/1012.2466}{\texttt{arXiv:1012.2466}}.

\bibitem{Kidonakis:2010dk}
\hrefCMSnoop {}{N.~Kidonakis, ``{Next-to-next-to-leading soft-gluon corrections
  for the top quark cross section and transverse momentum distribution}'',}
  \textit{ Phys. Rev. D} \textbf{ 82} (2010) 114030,
  \href{http://dx.doi.org/10.1103/PhysRevD.82.114030}{\doi{10.1103/PhysRevD.82.114030}},
\href{http://www.arXiv.org/abs/1009.4935}{\texttt{arXiv:1009.4935}}.

\bibitem{MCFM:diboson}
\hrefCMSnoop {}{J.~M. Campbell, R.~K. Ellis, and C.~Williams, ``{Vector boson
  pair production at the LHC}'',} \textit{ JHEP} \textbf{ 07} (2011) 018,
  \href{http://dx.doi.org/10.1007/JHEP07(2011)018}{\doi{10.1007/JHEP07(2011)018}},
\href{http://www.arXiv.org/abs/1105.0020}{\texttt{arXiv:1105.0020}}.

\bibitem{bib:CLs1}
\hrefCMSnoop {}{A.~L. Read, ``Presentation of search results: the {$CL_s$}
  technique'',} \textit{ J. Phys. G} \textbf{ 28} (2002) 2693,
\href{http://dx.doi.org/10.1088/0954-3899/28/10/313}{\doi{10.1088/0954-3899/28/10/313}}.

\bibitem{bib:CLs2}
\hrefCMSnoop {}{T.~Junk, ``{Confidence level computation for combining searches
  with small statistics}'',} \textit{ Nucl. Instrum. Meth. A} \textbf{ 434}
  (1999) 435,
  \href{http://dx.doi.org/10.1016/S0168-9002(99)00498-2}{\doi{10.1016/S0168-9002(99)00498-2}},
\href{http://www.arXiv.org/abs/hep-ex/9902006}{\texttt{arXiv:hep-ex/9902006}}.

\bibitem{bib:STAT_RooStats}
\href {http://pos.sissa.it/cgi-bin/reader/conf.cgi?confid=93}{L.~{Moneta},
  K.~{Cranmer}, G.~{Schott}, and W.~{Verkerke}, ``The {R}oo{S}tats
  {P}roject'',} in \textit{ Proceedings of the 13th International Workshop on
  Advanced Computing and Analysis Techniques in Physics Research (ACAT2010)}.
\newblock SISSA, 2010.
\newblock \href{http://www.arXiv.org/abs/1009.1003}{\texttt{arXiv:1009.1003}}.

\bibitem{bib:CT10}
H.-L. Lai\hrefCMSnoop {}{ {et~al.}, ``New parton distributions for collider
  physics'',} \textit{ Phys. Rev. D} \textbf{ 82} (2010) 074024,
  \href{http://dx.doi.org/10.1103/PhysRevD.82.074024}{\doi{10.1103/PhysRevD.82.074024}},
\href{http://www.arXiv.org/abs/1007.2241}{\texttt{arXiv:1007.2241}}.

\bibitem{bib:MSTW2008}
\hrefCMSnoop {}{A.~Martin, W.~Stirling, R.~Thorne, and G.~Watt, ``{Parton
  distributions for the LHC}'',} \textit{ Eur. Phys. J. C} \textbf{ 63} (2009)
  189,
  \href{http://dx.doi.org/10.1140/epjc/s10052-009-1072-5}{\doi{10.1140/epjc/s10052-009-1072-5}},
\href{http://www.arXiv.org/abs/0901.0002}{\texttt{arXiv:0901.0002}}.

\bibitem{bib:NNPDF}
\hrefCMSnoop {}{{NNPDF} Collaboration, ``A first unbiased global {NLO}
  determination of parton distributions and their uncertainties'',} \textit{
  Nucl. Phys. B} \textbf{ 838} (2010) 136,
  \href{http://dx.doi.org/10.1016/j.nuclphysb.2010.05.008}{\doi{10.1016/j.nuclphysb.2010.05.008}},
\href{http://www.arXiv.org/abs/1002.4407}{\texttt{arXiv:1002.4407}}.

\bibitem{bib:RoniLHC}
\hrefCMSnoop {}{P.~J. Fox, R.~Harnik, J.~Kopp, and Y.~Tsai, ``{Missing energy
  signatures of dark matter at the LHC}'',} \textit{ Phys. Rev. D} \textbf{ 85}
  (2012) 056011,
  \href{http://dx.doi.org/10.1103/PhysRevD.85.056011}{\doi{10.1103/PhysRevD.85.056011}},
\href{http://www.arXiv.org/abs/1109.4398}{\texttt{arXiv:1109.4398}}.

\bibitem{An:2012va}
\hrefCMSnoop {}{H.~An, X.~Ji, and L.-T. Wang, ``Light dark matter and {$Z'$}
  dark force at colliders'',} \textit{ JHEP} \textbf{ 07} (2012) 182,
  \href{http://dx.doi.org/10.1007/JHEP07(2012)182}{\doi{10.1007/JHEP07(2012)182}},
\href{http://www.arXiv.org/abs/1202.2894}{\texttt{arXiv:1202.2894}}.

\bibitem{Friedland:2011za}
\hrefCMSnoop {}{A.~Friedland, M.~L. Graesser, I.~M. Shoemaker, and L.~Vecchi,
  ``Probing nonstandard standard model backgrounds with {LHC} monojets'',}
  \textit{ Phys. Lett. B} \textbf{ 714} (2012) 267,
  \href{http://dx.doi.org/10.1016/j.physletb.2012.06.078}{\doi{10.1016/j.physletb.2012.06.078}},
\href{http://www.arXiv.org/abs/1111.5331}{\texttt{arXiv:1111.5331}}.

\bibitem{Buchmueller:2013dya}
\hrefCMSnoop {}{O.~Buchmueller, M.~J. Dolan, and C.~McCabe, ``Beyond effective
  field theory for dark matter searches at the {LHC}'',} \textit{ JHEP}
  \textbf{ 01} (2014) 025,
  \href{http://dx.doi.org/10.1007/JHEP01(2014)025}{\doi{10.1007/JHEP01(2014)025}},
\href{http://www.arXiv.org/abs/1308.6799}{\texttt{arXiv:1308.6799}}.

\bibitem{bib:COGENT}
\hrefCMSnoop {}{{CoGeNT} Collaboration, ``Results from a Search for Light-Mass
  Dark Matter with a $p$-Type Point Contact Germanium Detector'',} \textit{
  Phys. Rev. Lett.} \textbf{ 106} (2011) 131301,
  \href{http://dx.doi.org/10.1103/PhysRevLett.106.131301}{\doi{10.1103/PhysRevLett.106.131301}},
  \href{http://www.arXiv.org/abs/1002.4703}{\texttt{arXiv:1002.4703}}.

\bibitem{SIMPLE2012}
\hrefCMSnoop {}{{SIMPLE} Collaboration, ``Final Analysis and Results of the
  Phase {II} {SIMPLE} Dark Matter Search'',} \textit{ Phys. Rev. Lett.}
  \textbf{ 108} (2012) 201302,
  \href{http://dx.doi.org/10.1103/PhysRevLett.108.201302}{\doi{10.1103/PhysRevLett.108.201302}},
\href{http://www.arXiv.org/abs/1106.3014}{\texttt{arXiv:1106.3014}}.

\bibitem{bib:COUPP2012}
\hrefCMSnoop {}{{COUPP} Collaboration, ``First dark matter search results from
  a 4-kg {CF$_3$I} bubble chamber operated in a deep underground site'',}
  \textit{ Phys. Rev. D} \textbf{ 86} (2012) 052001,
  \href{http://dx.doi.org/10.1103/PhysRevD.86.052001}{\doi{10.1103/PhysRevD.86.052001}},
\href{http://www.arXiv.org/abs/1204.3094}{\texttt{arXiv:1204.3094}}.

\bibitem{bib:CDMSII2010}
\hrefCMSnoop {}{{CDMS-II} Collaboration, ``{Dark Matter Search Results from the
  CDMS II Experiment}'',} \textit{ Science} \textbf{ 327} (2010) 1619,
  \href{http://dx.doi.org/10.1126/science.1186112}{\doi{10.1126/science.1186112}},
\href{http://www.arXiv.org/abs/0912.3592}{\texttt{arXiv:0912.3592}}.

\bibitem{bib:CDMSII2011}
\hrefCMSnoop {}{{CDMS} Collaboration, ``Results from a Low-Energy Analysis of
  the {CDMS II} Germanium Data'',} \textit{ Phys. Rev. Lett.} \textbf{ 106}
  (2011) 131302,
  \href{http://dx.doi.org/10.1103/PhysRevLett.106.131302}{\doi{10.1103/PhysRevLett.106.131302}},
  \href{http://www.arXiv.org/abs/1011.2482}{\texttt{arXiv:1011.2482}}.

\bibitem{SuperCDMS}
\hrefCMSnoop {}{{SuperCDMS} Collaboration, ``{CDMSlite: A Search for Low-Mass
  WIMPs using Voltage-Assisted Calorimetric Ionization Detection in the
  SuperCDMS Experiment}'',} \textit{ Phys. Rev. Lett.} \textbf{ 112} (2014)
  041302,
  \href{http://dx.doi.org/10.1103/PhysRevLett.112.041302}{\doi{10.1103/PhysRevLett.112.041302}},
\href{http://www.arXiv.org/abs/1309.3259}{\texttt{arXiv:1309.3259}}.

\bibitem{bib:XENON100}
\hrefCMSnoop {}{{XENON100} Collaboration, ``Dark Matter Results from 100 Live
  Days of {XENON100} Data'',} \textit{ Phys. Rev. Lett.} \textbf{ 107} (2011)
  131302,
  \href{http://dx.doi.org/10.1103/PhysRevLett.107.131302}{\doi{10.1103/PhysRevLett.107.131302}},
  \href{http://www.arXiv.org/abs/1104.2549}{\texttt{arXiv:1104.2549}}.

\bibitem{bib:LUX}
\hrefCMSnoop {}{{LUX} Collaboration, ``First results from the {LUX} dark matter
  experiment at the {Sanford Underground Research Facility}'',} \textit{ Phys.
  Rev. Lett.} \textbf{ 112} (2014) 091303,
  \href{http://dx.doi.org/10.1103/PhysRevLett.112.091303}{\doi{10.1103/PhysRevLett.112.091303}},
\href{http://www.arXiv.org/abs/1310.8214}{\texttt{arXiv:1310.8214}}.

\bibitem{bib:CDMSSi}
\hrefCMSnoop {}{{CDMS} Collaboration, ``{Silicon Detector Dark Matter Results
  from the Final Exposure of CDMS II}'',} \textit{ Phys. Rev. Lett.} \textbf{
  111} (2013) 251301,
  \href{http://dx.doi.org/10.1103/PhysRevLett.111.251301}{\doi{10.1103/PhysRevLett.111.251301}},
\href{http://www.arXiv.org/abs/1304.4279}{\texttt{arXiv:1304.4279}}.

\bibitem{SUPERK}
\hrefCMSnoop {}{{Super-Kamiokande} Collaboration, ``An indirect search for
  {WIMPs} in the sun using 3109.6 days of upward-going muons in
  {Super-Kamiokande}'',} \textit{ Astrophys. J.} \textbf{ 742} (2011) 78,
  \href{http://dx.doi.org/10.1088/0004-637X/742/2/78}{\doi{10.1088/0004-637X/742/2/78}},
\href{http://www.arXiv.org/abs/1108.3384}{\texttt{arXiv:1108.3384}}.

\bibitem{IceCube:2011aj}
\hrefCMSnoop {}{{IceCube} Collaboration, ``{Multi-year search for dark matter
  annihilations in the Sun with the AMANDA-II and IceCube detectors}'',}
  \textit{ Phys. Rev. D} \textbf{ 85} (2012) 042002,
  \href{http://dx.doi.org/10.1103/PhysRevD.85.042002}{\doi{10.1103/PhysRevD.85.042002}},
\href{http://www.arXiv.org/abs/1112.1840}{\texttt{arXiv:1112.1840}}.

\bibitem{Karg:2009xk}
\hrefCMSnoop {}{S.~Karg, M.~Kr{\"a}mer, Q.~Li, and D.~Zeppenfeld, ``{NLO QCD
  corrections to graviton production at hadron colliders}'',} \textit{ Phys.
  Rev. D} \textbf{ 81} (2010) 094036,
  \href{http://dx.doi.org/10.1103/PhysRevD.81.094036}{\doi{10.1103/PhysRevD.81.094036}},
\href{http://www.arXiv.org/abs/0911.5095}{\texttt{arXiv:0911.5095}}.

\bibitem{bib:DELPHI}
\hrefCMSnoop {}{{DELPHI} Collaboration, ``Photon events with missing energy in
  e$^{+}$e$^{-}$ collisions at $\sqrt{s} = 130$ to 209 {GeV}'',} \textit{ Eur.
  Phys. J. C} \textbf{ 38} (2005) 395,
  \href{http://dx.doi.org/10.1140/epjc/s2004-02051-8}{\doi{10.1140/epjc/s2004-02051-8}},
\href{http://www.arXiv.org/abs/hep-ex/0406019}{\texttt{arXiv:hep-ex/0406019}}.

\bibitem{bib:Kathrein}
\hrefCMSnoop {}{S.~Kathrein, S.~Knapen, and M.~J. Strassler, ``Bounds from
  {LEP} on unparticle interactions with electroweak bosons'',} \textit{ Phys.
  Rev. D} \textbf{ 84} (2011) 015010,
  \href{http://dx.doi.org/10.1103/PhysRevD.84.015010}{\doi{10.1103/PhysRevD.84.015010}},
\href{http://www.arXiv.org/abs/1012.3737}{\texttt{arXiv:1012.3737}}.

\end{thebibliography}\endgroup

\cleardoublepage \appendix\section{The CMS Collaboration \label{app:collab}}\begin{sloppypar}\hyphenpenalty=5000\widowpenalty=500\clubpenalty=5000\textbf{Yerevan Physics Institute,  Yerevan,  Armenia}\\*[0pt]
V.~Khachatryan, A.M.~Sirunyan, A.~Tumasyan
\vskip\cmsinstskip
\textbf{Institut f\"{u}r Hochenergiephysik der OeAW,  Wien,  Austria}\\*[0pt]
W.~Adam, T.~Bergauer, M.~Dragicevic, J.~Er\"{o}, C.~Fabjan\cmsAuthorMark{1}, M.~Friedl, R.~Fr\"{u}hwirth\cmsAuthorMark{1}, V.M.~Ghete, C.~Hartl, N.~H\"{o}rmann, J.~Hrubec, M.~Jeitler\cmsAuthorMark{1}, W.~Kiesenhofer, V.~Kn\"{u}nz, M.~Krammer\cmsAuthorMark{1}, I.~Kr\"{a}tschmer, D.~Liko, I.~Mikulec, D.~Rabady\cmsAuthorMark{2}, B.~Rahbaran, H.~Rohringer, R.~Sch\"{o}fbeck, J.~Strauss, A.~Taurok, W.~Treberer-Treberspurg, W.~Waltenberger, C.-E.~Wulz\cmsAuthorMark{1}
% ...
\vskip\cmsinstskip
% ...
1:~~Also at Vienna University of Technology, Vienna, Austria\\
2:~~Also at CERN, European Organization for Nuclear Research, Geneva, Switzerland\\
% ...

\end{sloppypar}
\end{document}